\documentclass[fleqn,usenatbib]{mnras}

\usepackage{newtxtext,newtxmath}

\usepackage[T1]{fontenc}

\DeclareRobustCommand{\VAN}[3]{#2}
\let\VANthebibliography\thebibliography
\def\thebibliography{\DeclareRobustCommand{\VAN}[3]{##3}\VANthebibliography}

\usepackage{graphicx}	
\usepackage{amsmath}	
\usepackage[pdftex,dvipsnames]{xcolor}
\usepackage{multirow}
\usepackage{array, makecell}

\newcommand{\cosmos}{COSMOS2015}
\DeclareMathOperator\erf{erf}
\definecolor{myorange}{RGB}{255, 104, 51}

\newcommand{\angstrom}{\text{\normalfont\AA}}
\newcommand{\OII}{\mathrm{OII}}
\newcommand{\OIII}{\mathrm{OIII}}
\newcommand{\Halpha}{\mathrm{H}_{\alpha}}
\newcommand{\Hbeta}{\mathrm{H}_{\beta}}
\newcommand{\iab}{i_{\mathrm{AB}}}
\newcommand{\MNUV}{\mathrm{M}_\mathrm{NUV}}
\newcommand{\Lya}{\mathrm{Ly}_{\alpha}}

\newcommand{\PAUSdatareduction}{Serrano in preparation}
\newcommand{\PAUSphotocalibration}{Castander in preparation}
\newcommand{\PAUSforcedphotometry}{Gaztanaga in preparation}

\title[PAUS+COSMOS photo-$z$ sample]{The PAU Survey: An improved photo-$z$ sample in the COSMOS field}

\author[A. Alarcon et al.]{Alex Alarcon$^{1,2,3}$\thanks{E-mail: alexalarcongonzalez@gmail.com (AA)},
Enrique Gaztanaga$^{2,3}$,
Martin Eriksen$^{4}$\thanks{Also at Port d'Informaci\'{o} Cient\'{i}fica (PIC), Campus UAB, C. Albareda s/n, 08193 Bellaterra (Cerdanyola del Vall\`{e}s), Spain},
Carlton M. Baugh$^{5,6}$,
\newauthor
Laura Cabayol$^{4}$,
Ricard Casas$^{2,3}$,
Jorge Carretero$^{4}\footnotemark[2]$,
Francisco J. Castander$^{2,3}$,
\newauthor
Juan De Vicente$^{7}$, 
Enrique Fernandez$^{4}$, 
Juan Garcia-Bellido$^{8}$,
Hendrik Hildebrandt$^{9}$,
\newauthor
Henk Hoekstra$^{10}$,
Benjamin Joachimi$^{11}$,
Giorgio Manzoni$^{5,6,12}$,
Ramon Miquel$^{4,13}$,
\newauthor
Peder Norberg$^{5,6,12}$,
Cristobal Padilla$^{4}$,
Pablo Renard$^{2,3}$,
Eusebio Sanchez$^{7}$,
\newauthor
Santiago Serrano$^{2,3}$,
Ignacio Sevilla-Noarbe$^{7}$,
Malgorzata Siudek$^{4}$,
Pau Tallada-Cresp{\'i}$^{7\footnotemark[2]}$
\\
\\
\emph{\it\normalsize (Author affiliations are listed after the references)}
}

\pubyear{2020}

\begin{document}
\label{firstpage}
\pagerange{\pageref{firstpage}--\pageref{lastpage}}
\maketitle

\begin{abstract}
We present -- and make publicly available -- accurate and precise photometric redshifts in the ACS footprint from the COSMOS field for objects with $i_{\mathrm{AB}}\leq 23$. The redshifts are computed using a combination of narrow band photometry from PAUS, a survey with 40 narrow bands spaced at $100\angstrom$ intervals covering the range from $4500\angstrom$ to $8500\angstrom$, and 26 broad, intermediate, and narrow bands  covering the UV, visible and near infrared spectrum from the COSMOS2015 catalogue. We introduce a new method that models the spectral energy distributions (SEDs) as a linear combination of continuum and emission line templates and computes its Bayes evidence, integrating over the linear combinations. The correlation between the UV luminosity and the OII line is measured using the 66 available bands with the zCOSMOS spectroscopic sample, and used as a prior which constrains the relative flux between continuum and emission line templates. The flux ratios between the OII line and $\Halpha$, $\Hbeta$ and $\OIII$ are similarly measured and used to generate the emission line templates. Comparing to public spectroscopic surveys via the quantity $\Delta_z\equiv(z_{\mathrm{photo}}-z_{\mathrm{spec}})/(1+z_{\mathrm{spec}})$, we find the photometric redshifts to be more precise than previous estimates, with $\sigma_{68}(\Delta_z) \approx (0.003, 0.009)$ for galaxies at magnitude  $i_{\mathrm{AB}}\sim18$ and $i_{\mathrm{AB}}\sim23$, respectively, which is $3\times$ and $1.66\times$ tighter than COSMOS2015. Additionally, we find the redshifts to be very accurate on average, yielding a median of the $\Delta_z$ distribution compatible with $|\mathrm{median}(\Delta_z)|\leq0.001$ at all redshifts and magnitudes considered. 
Both the added PAUS data and new methodology contribute significantly to the improved results.
The catalogue produced with the technique presented here is expected to provide a robust redshift calibration for current and future lensing surveys, and allows one to probe galaxy formation physics in an unexplored luminosity-redshift regime, thanks to its combination of depth, completeness and excellent redshift precision and accuracy.
\end{abstract}

\begin{keywords}
galaxies: distances and redshifts -- galaxies: evolution -- galaxies: photometry
\end{keywords}



\section{Introduction}

Redshift galaxy surveys can be broadly divided into two categories: spectroscopic surveys and imaging surveys. The former obtains a high resolution spectra of the object within some wavelength coverage, which is used to identify sharp features like emission and absorption lines to nail the redshift of the object with very high precision. However, these are expensive to obtain: they require knowing the position of the object beforehand and a large exposure time, which makes it observationally inefficient to observe faint objects over a large area. Such surveys also suffer from incompleteness both because not all objects in the field are always targeted and also since a fraction of the measured spectra fail to provide an accurate redshift, for example for the  lack of obvious emission or absorption features in low signal-to-noise spectra, when only one line is observed, or when there is a line confusion. In contrast, imaging surveys are able to obtain measurements of every object in the field of view at the same time from a set of bandpass filtered images, which allows to cover large areas faster and to a greater depth. This happens at the expense of getting flux measurements with very poor spectral resolution since the width of typical broad band filters is larger than 100nm, which makes the photometric redshift (i.e. photo-$z$) determination much less precise and sometimes inaccurate. 

The precision of photometric redshift surveys can be improved with narrower bands and with a broader wavelength range coverage \citep[\textit{e.g.} see Figure 3 from ][]{Salvato19}, and in recent years a new generation of multi-band photometric imaging surveys have emerged, for example, spanning from the ultraviolet to infrared \citep[COSMOS, ][]{COSMOSILBERT}, using sets of intermediate bands \citep[SHARDS or ALHAMBRA, ][]{shards13, alhambra}, or a number of narrow band filters \citep[PAUS, ][]{PAUSphotoz}. From the more precise photometric redshifts one can extract better measurements of galaxy properties (luminosity, stellar mass, star formation rate) to probe and understand galaxy formation and galaxy evolution physics with a denser sample, at higher redshifts, and with little selection effects. One can measure galaxy clustering in thin redshift shells as a function of these intrinsic galaxy properties, measure luminosity functions and star formation histories, or use  galaxy-galaxy lensing to constrain the galaxy to halo connection in an less explored luminosity-redshift regime. In particular, the Physics of the Accelerating Universe Survey (PAUS) is an ongoing narrow band imaging survey that intends to cover $100\mathrm{deg}^2$ using 40 narrow band filters, increasing the number of objects with subpercent redshift precision by two orders of magnitude \citep{PAUSphotoz}. The role of environment in structure formation is limited by the poor redshift precision in broad band surveys and by the tiny area or low density in spectroscopic surveys. The unique combination of area, depth and redshift resolution from PAUS allows to sample with high density several galaxy populations, with which studies targeting nonlinear galaxy bias, intrinsic alignments, magnification or density field reconstruction can be developed.

On the other hand, imaging weak lensing galaxy surveys have entered the era of precision cosmology and have become one of the most powerful probes for the $\Lambda$CDM cosmological model by measuring the shape and position of hundreds of millions of galaxies. Current and future surveys such as the Dark Energy Survey \citep[DES,][]{DES_troxel_19,DES_3x2pt,2018ApJS..239...18A}, the Kilo-Degree Survey  \citep[KiDS,][]{Hildebrandt2017,KIDS_cosmicshear_Hilde_20,KIDS_dataset_Wright_19,KIDS_cosmology_Wright_20}, Hyper Suprime-Cam survey \citep[HSC,][]{HSC_DR1_aihara_18,HSC_hikage_19}, the Legacy Survey of Space and Time \citep[LSST,][]{LSST}, or the \textit{Euclid} mission \citep{Euclid} are reaching or will reach a point where systematic uncertainties limit the full exploitation of their statistical power. Among these, one of the most challenging systematic uncertainties is the characterisation of the redshift distribution of the weak lensing tomographic samples, which contain millions of faint galaxies with a few colours measured using broad band filters. A correct description of such redshift distributions is crucial to avoid introducing a bias in the cosmological inference \citep{Huterer2006,cfht_hilde,Cunha2012,Benjamin2013,Huterer2013,Bonnett2015,Joudaki_2017_cfht,Hoyle2017,Hildebrandt2017,KidsDEScomb} and to allow a robust comparison between cosmological parameters from weak lensing analysis and from the cosmic microwave background \citep[CMB,][]{planck18}, especially when a number of recent studies suggest a mild tension between the values of cosmological parameters inferred for the early and late time universe \citep{KidsDEScomb,Asgari19,KIDS_cosmology_Wright_20}.

The really large area and depth covered by these surveys makes it unfeasible to measure spectroscopic redshifts for each galaxy of interest, which is why a number of alternative techniques have been developed over the years to estimate their redshift distributions. These can be broadly grouped as those
 which use angular cross-correlations with an overlapping tracer sample with well characterised redshifts \citep[\textit{clustering redshifts}, see][]{Newman2008,Menard2013,Schmidt2013,Gatti2018,Davis2017,Hildebrandt2017}, those that model the galaxy spectral energy distribution (SED) of each galaxy to connect their observed colours to redshift \citep[e.g.][]{BPZ,lephare,mizuki_tanaka15,Hoyle2017} and those that model the colour redshift relation empirically using calibration samples \citep[e.g.][]{Cunha2012,dnf,Bonnett2015,Buchs2019,Wright19KidsSOM}. Each of these methods present different intrinsic systematics and potential biases, and are typically best used in combination \citep[e.g.][]{Hildebrandt2017,Hoyle2017}.

Direct or empirical calibration methods rely on galaxy samples where abundant redshift information is available, either through spectroscopy or many band photometric redshifts. The former can be very accurate, but estimates are only available for a subset of the sample, and the targeting strategy, quality selection and incompleteness can introduce a statistical redshift bias with respect to the redshift of the full sample \citep{Bonnett2015,Gruen17,Speagle2019,Hartley_20,KIDS_cosmology_Wright_20}. On the other hand, multi-band photometric surveys provide a complete redshift sample at the expense of degrading the redshift precision, and can be biased if the galaxy SED modelling is incorrect. The COSMOS field \citep{Scoville2007_COSMOS} contains the most widely used multiband redshift calibration survey, which provides a unique combination of deep photometric observations ranging from the UV to the infrared over an area of $\sim2\deg^2$, and several photometric redshift catalogues have been produced over the years \citep{COSMOSILBERT,Ilbert13,COSMOS2015}.

Here, we combine the multiband photometry from \cite{COSMOS2015}, hereafter \cosmos,  with 40 narrow band filters from the PAUS survey \citep{PAUCampaper} which span the wavelength range from $4500\angstrom$ to $8500\angstrom$. The unique PAUS photometric set is able to determine very precise photometric redshifts thanks to its exquisite wavelength sampling, which specifies precisely the location of the very sharp features in the galaxy SED \citep{PAUSphotoz,deepz}. To estimate the photometric redshifts we develop an algorithm that models the galaxy SED as a linear combination of continuum and emission line templates, and marginalises over different combinations computing a Bayesian integral. Furthermore, we calibrate priors between the continuum and emission line templates using a subsample with spectroscopic redshifts and the multi-band photometry. In \cite{PAUSphotoz} a similar model (\textsc{bcnz2}) was used, where the best fitting linear combination of templates was calculated for each galaxy and model instead of the Bayesian integral. There, \textsc{bcnz2} was used to measure redshifts using the 40 narrow bands from PAUS and a subset of 6 broad bands from \cosmos, for objects with $\iab\leq22.5$ until redshift $z_{\mathrm{max}}=1.2$. Here, we use a total of 66 bands which include 40 narrow bands from PAUS and a combination of 26 narrow, intermediate, and broad bands from \cosmos, and we extend the magnitude and redshift limits to $\iab\leq23$ and $z_{\mathrm{max}}=3$. We focus on objects that were identified as galaxies in \cosmos, removing AGN X-ray detected sources, for which several photo-$z$ studies exist \citep{Salvato2009,Salvato2011,Simm2015,Marchesi2016}. We make the redshift catalogue publicly available, including the redshift distribution of each object.

This paper is organised as follows. In section~\ref{sec:data} we describe the photometric data catalogues and the spectroscopic redshift catalogue used in this work. Section~\ref{sec:methodology} describes the methodology used to describe the galaxy SED and infer the photometric redshift. Section~\ref{sec:results} presents the primary photometric redshift results of this work. In section~\ref{sec:discussion} we discuss more details of the analysis and possible extensions for future work. We conclude and summarise in section~\ref{sec:conclusions}. There are five appendices, which contain details of how to download the catalogue (Appendix~\ref{app:cosmohub}), details of how we combine heterogeneous data (Appendix~\ref{NB2BB}), details of the photo-$z$ algorithms (Appendix~\ref{algorithm}), the calibration of zero point offsets (Appendix~\ref{app:sysoffs}), and the population prior on the models (Appendix~\ref{app:popprior}).

\section{Data} \label{sec:data}

In this section we describe the data we are going to use throughout. We will use narrow band photometry from the PAU Survey, a combination of narrow, intermediate and broad bands coming from various instruments publicly released by the COSMOS Survey, and a spectroscopic redshift catalogue including measurements from several public redshift surveys.

\subsection{PAUCam narrow band photometry}

The Physics of the Accelerating Universe Survey (PAUS) is an ongoing imaging survey using a unique instrument, PAUCam (\citealt{PAUCampaper}), mounted at the William Herschel Telescope (WHT) and located in the Observatorio del Roque de los Muchachos (La Palma, Canary Islands, Spain). PAUCam carries a set of 40 narrow band (NB) filters with 12.5nm FWHM that span the wavelength range from 450nm to 850nm, in steps of 100nm. PAUS has been collecting data since 2015 during several observing runs imaging five different fields: the COSMOS field \citep{Scoville2007_COSMOS} and the CFHT W1, W2, W3 and W4 fields\footnote{\url{http://www.cfht.hawaii.edu/Science/CFHLS/cfhtlsdeepwidefields.html}}. The PAU/CFHT fields are larger and represent the main survey, which is intended to observe up to $100\deg^2$, while the COSMOS field (2 $\deg^2$) has been targeted as a calibration field since many photometric observations already exist, ranging from ultraviolet all the way to far infrared, as well as spectroscopic surveys with relatively high completeness and depth. In this work we use the data collected in the COSMOS field from campaigns between 2015 and 2017.

\subsubsection{Data reduction overview}

At the end of each observing night, the data taken at WHT is sent to Port d'Informació Científica (PIC) for its storage and processing (\citealt{PAUSoperation}). The data reduction process starts with initial de-trending, where a number of signatures from the instrument are removed from the images, using the \textsc{nightly} pipeline (see \PAUSdatareduction, \PAUSphotocalibration~for details). This includes removing electronic bias with an overscan subtraction, correcting the gain from the different amplifiers and compensation from readout patterns using bias frames. A master flat is created from exposures of the dome with homogeneous illumination that are taken every afternoon before the observation. It is used to correct the vignetting of the telescope corrector, among other effects such as dead and hot pixels. Each individual narrow band filter is only covering a single CCD, instead of a unique broad band that covers all the focal plane, and the visible edges of the filters in the supporting grid of the filter tray produced scattered light in the image edges. An adjustment to the camera in 2016 significantly reduced this effect, which is partly mitigated by the pipeline by using a low pass filter with sigma clipping. Cosmic rays are identified using a Laplacian edge detection (\citealt{laplaciancosmicrays}) and masked from the image.

An astrometric solution is added to align the different exposures using \textsc{scamp} (\citealt{PSFEX}) by comparing to \textit{GAIA} DR1 (\citealt{GaiaDR1}). To estimate the Point Spread Function (PSF) and perform the photometric narrow band calibration, we select stars with $i_{\mathrm{AB}}<21$ from the Sloan Digital Sky Survey (SDSS)  \citep{SDSSstandardsystem}. The PSF is modelled using \textsc{psfex} (\citealt{PSFEX}) in these stars. We calibrate the narrow band photometry (\PAUSphotocalibration) using the SDSS $u$, $g$, $r$, $i$, and $z$  \texttt{psfMag}  magnitudes of these stars, performing the following procedure. We fit the Pickles stellar library \citep{Pickles98} to the SDSS photometry of the star assuming the Milky Way extinction of the star is given by the value from the Planck 2015 thermal dust map \citet{Planckdustmap2016}. Here we approximate all stars to go through all the dust of the MW in that direction, since the majority of our calibration stars are fainter than magnitude 17-18 in the $r$-band. Then, we generate synthetic narrow band observations without the Milky Way extinction for each SED and combine them with a weight equal to their probability from the SDSS fit. This combined synthetic narrow band is compared to observations to obtain the narrow band zero points for the star. Finally, all star zero points in the same image are combined to obtain one zero point per image and narrow band. Note that this procedure also corrects for Milky Way extinction as the synthetic narrow band fluxes are generated excluding the Milky Way extinction.


Narrow band photometry is obtained using the \textsc{memba} pipeline (Multi-Epoch and Multi-Band Analysis, \PAUSdatareduction, \PAUSforcedphotometry). In general, we rely on deep overlapping observations from lensing surveys to provide a detection catalogue with high quality shape measurements to perform forced aperture photometry. In the COSMOS field, positions and shape measurements from ACS are used. The half light radius, $r_{50}$, is used along with ellipticity measurements from \cite{morphizurich} and the PAUS PSF FWHM to determine the aperture size and shape to target $62.5\%$ of the light, set to optimize the signal to noise. A flux measurement is obtained for each individual exposure using this aperture measurement and a background subtraction estimated from a fixed annulus of 30 to 45 pixels around the source, where sources falling in the annulus get sigma clipped \citep[for more details of the annulus background subtraction see][]{Cabayol19}. Fluxes measured in different exposures are corrected with the estimated image zeropoints and get combined with a weighted average to produce a narrow band coadded flux measurement. 

The data reduction pipeline propagates flags for each individual exposure and object, and flagged measurements (indicating problems in the photometry) are not included in the weighted average. We remove objects with fewer than 30 narrow band measurements.

\subsection{COSMOS survey photometry}

Along with the narrow band data described in the previous section, we include photometry from several filters from the released \cosmos~catalog\footnote{\url{ftp://ftp.iap.fr/pub/from_users/hjmcc/COSMOS2015/}} \citep{COSMOS2015}, with filters from ultraviolet to near infrared. Here we list the bands we use in this work, along with the original instrument/survey: NUV data from GALEX; $u^{*}$ from the Canada-France Hawaii Telescope (MegaCam); ($B$, $V$, $r$, $i^{+}$, $z^{++}$) broad bands, ($IA427$, $IA464$, $IA484$, $IA505$, $IA527$, $IA574$, $IA624$, $IA679$, $IA709$, $IA738$, $IA767$, $IA827$) intermediate bands and ($NB711$, $NB816$) narrow bands from Suprime-Cam/Subaru; $Y$ broad band from HSC/Subaru; ($Y$, $J$, $H$, $K_s$) from VIRCAM/VISTA (UltraVISTA-DR2); and ($H$, $K_s$) data from WIRCam/CFHT. 

We point the reader to \citep{COSMOS2015} and references therein for a detailed overview of these observations and the data reduction. We use the $3\arcsec$ diameter PSF homogenised coadded flux measurements available in \cosmos~ and apply several corrections as described and provided in the catalogue release.
In particular, we correct for Milky Way dust extinction using the $E(B-V)$ value available for each galaxy and an effective factor $F_{x}$ for each filter $x$ \citep[see Table 3 and Equation 10 in][]{COSMOS2015} using 
\begin{equation}
\mathrm{mag}_\mathrm{corrected} = \mathrm{mag}_\mathrm{uncorrected} - E(B-V)*F_x,
\end{equation}
where the factors $F_x$ are derived from the filter
response function and integrated against the galactic extinction curve (for a sense of scale, $F_{\mathrm{NUV}}\approx 8.6$ and $F_{z^{++}}\approx 1.5$). The fluxes have also been corrected from aperture to total flux, using the column \texttt{OFFSET}. In addition, we remove masked objects by using the flag parameter \texttt{FLAG\_PETER=0}, we remove both objects identified as stars or AGN by selecting  \texttt{TYPE=0}, and we select objects with \texttt{0<PHOTOZ<9}. 

\subsection{Combined photometric catalogue}\label{sec:combined_catalog}

We combine the narrow band catalogue from PAUS and the photometry from \cosmos~by matching objects by (ra, dec) position, keeping objects within $1\arcsec$ radius. We restrict the analysis to objects with an \textsc{AUTO} $i$ band magnitude brighter than $\iab\leq23$, which we obtain from \citet{COSMOSILBERT} ($\iab$ is also the reference magnitude in the PAUS data reduction). For objects fainter than this magnitude the PAUS narrow band photometry has a typical signal to noise well below 5 in all bands \citep[see Figure 2 in][]{PAUSphotoz}, but emission lines can still be significantly detected in the narrow band filters. We defer to future work how including PAUS NBs can improve the photo-z performance in this fainter regime. The final catalogue after masking contains \textsc{40672} galaxies within the ACS footprint \cite[see Figure 1 from][for different COSMOS footprints]{COSMOS2015}. 

The combination of heterogeneous photometry from different instruments can be complicated and ultimately degrade the photometric redshift performance of the catalogue if it is not performed consistently. While in the case of PAUS the flux measurement is obtained from a variable aperture that targets $62.5\%$ of the total light of each galaxy, adapting to the galaxy's apparent size and taking into account the PSF of each individual image, the \cosmos~photometry measures flux with a fixed aperture of $3\arcsec$ on PSF homogenised images, which is then corrected from aperture flux to total flux. Therefore, each survey measures a different fraction of the light for each galaxy, which depends on its apparent size. To deal with this effect, we have developed a self calibration algorithm that benefits from the overlap between the Subaru $r$-band and the PAUS narrow bands. We introduce a synthetic Subaru $r$-band flux, defined as a linear combination of narrow band flux measurements, and find a rescaling factor for each object that we apply to its narrow band photometry. In this way we homogenise the photometry across surveys. We describe this method in Appendix~\ref{NB2BB}. Finally, we note that the colours measured by each survey could also be slightly different, but we will consider this is a negligible effect.

In addition, we add a systematic error to each band for all objects, which accounts for any residual error that remains either from inaccuracies in the galaxy modelling, errors in the data reduction or from joining inconsistent heterogeneous photometry from different surveys. We add a $2\%$ error to every band except for the UltraVista broad bands $Y$, $J$, $H$, $K_s$ ($5\%$ error) and for the GALEX NUV band ($10\%$ error), similar to those used by \cosmos~(priv. comm.).

\begin{figure}
	\includegraphics[width=\columnwidth]{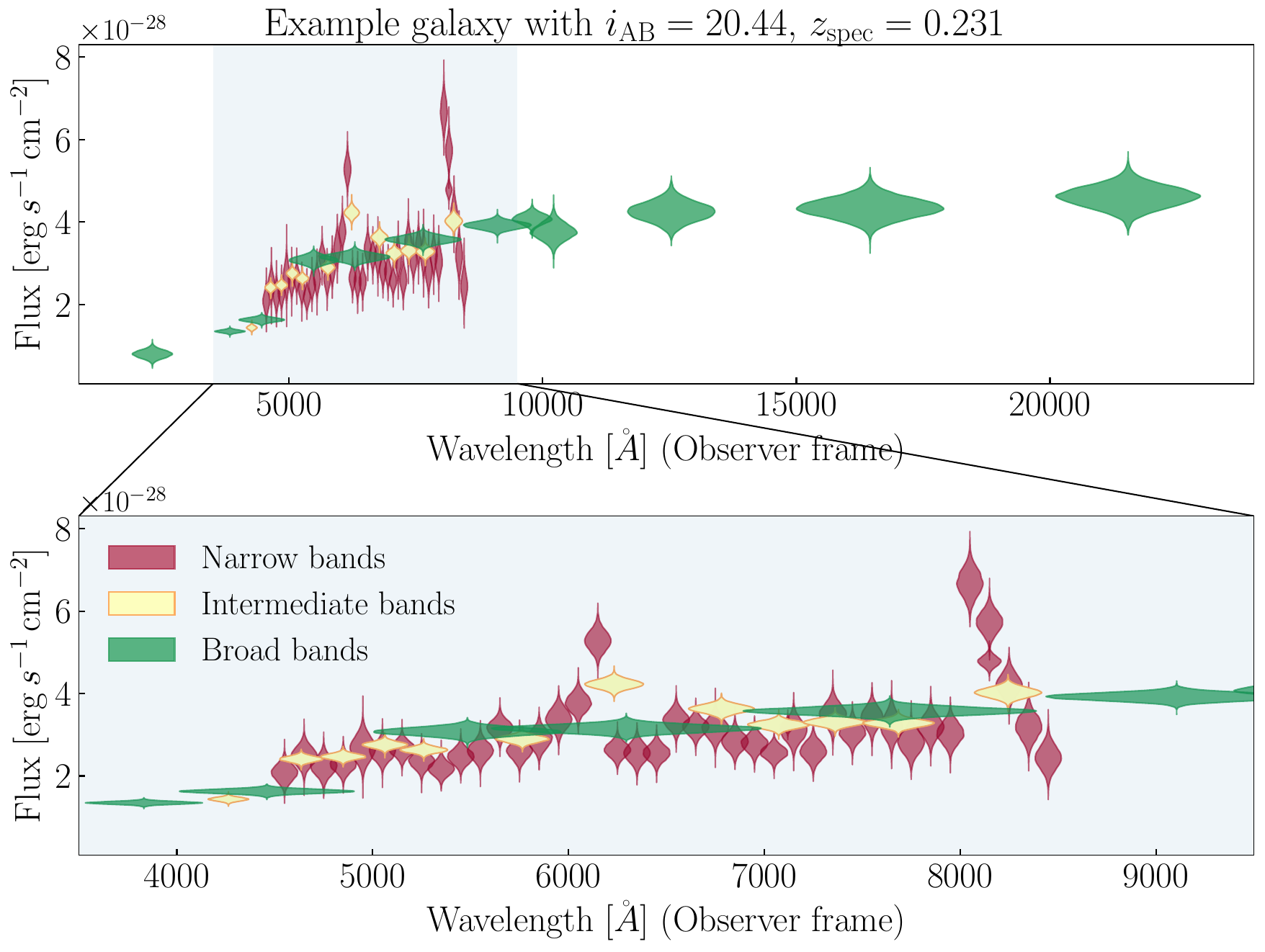}
    \caption{The 66 photometric bands used in this work for an example galaxy from the catalogue which has magnitude $\iab=20.44$ and spectroscopic redshift $z_{\mathrm{spec}}=0.231$. Bands with qualitatively similar FWHM have the same colour: broad bands in green, intermediate bands in yellow and narrow bands in red. The horizontal width of the violin plots shows the FWHM of each filter, while the violin plot is a Gaussian distribution centered at the measured flux and with variance equal to the measured flux variance plus an extra systematic variance. Two blocks of emission lines are clearly visible in the narrow bands, one containing the redshifted $\Halpha$ line around $\lambda\sim8100\angstrom$, and another containing the $\OIII$ doublet and $\Hbeta$ around $\lambda\sim6100\angstrom$.}
    \label{fig:example_galaxy}
\end{figure}

Figure~\ref{fig:example_galaxy} shows an example galaxy with the 66 photometric bands we use in this work, coloured according to their width (broad bands in green, intermediate bands in yellow and narrow bands in red). The horizontal width of the violins show the FWHM of each filter, while the violin is showing a Gaussian distribution centered at the measured flux and with variance equal to the measured flux variance plus the previously described extra systematic variance. Regarding the photometric band completeness, after flagging, 68\% of the catalogue has measurements available for all 66 bands, while 92\% of objects have at least 65 bands measured. All objects have measurements for the 12 intermediate bands and the four $Y$, $J$, $H$, $K_s$ UltraVista NIR bands. All bands have measurements available in at least 98\% of the catalogue, except for the GALEX NUV band which is missing in 23\% of the objects.

We make the photometric redshift catalogue publicly available along with the redshift distribution of each object (for details of the catalogue and how to download it see Appendix~\ref{app:cosmohub}).

\subsection{Spectroscopic data}
\label{sec:spectroscopic_data}

\begin{figure}
	\includegraphics[width=\columnwidth]{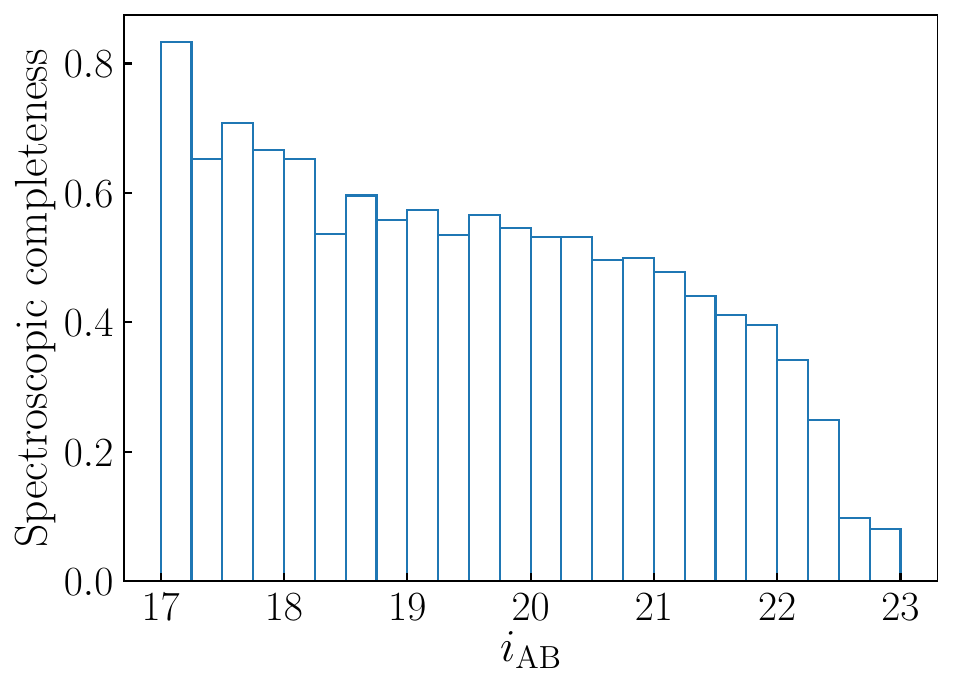}
    \caption{The spectroscopic completeness of the spectroscopic catalogue used in this work as a function of $\iab$ magnitude. The completeness is defined as the fraction of objects in the catalogue which have a spectroscopic redshift measured. The $\iab$ magnitude is the reference magnitude in the PAUS data reduction, and is the AUTO $i$ band magnitude from the \citet{COSMOSILBERT} catalogue.}
    \label{fig:spec_compl}
\end{figure}

To measure the precision and accuracy of the photometric redshifts we compare to spectroscopic redshifts. We use a compilation of public spectroscopic surveys (courtesy of Mara Salvato, private communication) and we apply a quality flag to keep only objects with a very reliable measurement. This compilation includes redshifts from the following instruments or surveys: zCOSMOS DR3 \citep{Lilly09zcosmos}, C3R2 DR1\&DR2 \cite{C3R2_DR1,C3R2_DR2}, 2dF \citep{2dF}, DEIMOS \citep{Casey17,Hasinger18,C3R2_DR2}, FMOS \citep{Kashino19}, LRIS \citep{Lee18}, MOSFIRE \citep{Kriek15}, MUSE \citep{Rosani19}, Magellan \citep{Calabro18}, VIS3COS \citep{Paulino18}. Whenever more than one redshift measurement is available for the same object we take the mean of all observations, and if the multiple observations disagree by more than 0.002 in redshift, we do not assign any spectroscopic redshift to that object (which removes 95 objects in the spectroscopic catalogue). The photometric catalogue from section~\ref{sec:combined_catalog} has a total of 12112 objects with an overlapping spectroscopic match. Figure~\ref{fig:spec_compl} shows a summary of the spectroscopic completeness of the catalogue as a function of $\iab$ magnitude, which stays above 40\% for bright magnitudes and decreases below 10\% for the fainter objects considered, at $\iab>22.5$. Since the spectroscopic redshifts in this catalog come from high resolution instruments, we expect their precision error to be negligible compared to the photo-$z$ precision. However, there could be a sample of outlier spectroscopic measurements that we cannot flag, since the amount of duplicate spectroscopic measurements is very small.

\section{Methodology} \label{sec:methodology}

In this section we describe the methodology used to model the spectral energy distribution of a galaxy and obtain its redshift distribution.

\subsection{Redshift distribution}

We want to obtain the redshift probability distribution of a galaxy $p(z|\pmb{f})$ from some flux observations $\pmb{f}$. To model the relation between redshift and $\pmb{f}$ we introduce a set of models denoted by $\{M\}$ that can predict the fluxes $\pmb{f}$ as a function of redshift for different type of galaxies. Therefore, we write
\begin{equation} \label{redshift_posterior}
p(z|\pmb{f}) = \sum_{\{M\}} p(z,M|\pmb{f}) \propto \sum_{\{M\}} p(\pmb{f}|z,M) p(z,M).
\end{equation}
Each model $M$ is defined as a linear combination with parameters $\{\alpha_j\}$ of a particular set of spectral energy distributions $\{t\}$ (SEDs), 
\begin{equation} \label{linear_model}
M(z) = \sum_{j} \alpha_j(z) \, t_{j}(z).
\end{equation}
The SED templates $\{t\}$ can either be continuum templates of different galaxy populations or the flux from emission lines. In this work we will use two continuum templates and one emission line template, so $\pmb{\alpha} = \{\alpha_{0}^{\mathrm{Cont}},\alpha_{1}^{\mathrm{Cont}}, \alpha^{\mathrm{EL}}\}$. For details on the SED templates see the next subsections. We predict the colours at different redshifts by redshifting the restframe SEDs and convolving with each filter accordingly. The amplitudes $\pmb{\alpha}$ are the free  parameters of the model at each redshift. The probability $p(\pmb{f}|z,M)$ is given by
\begin{equation} \label{bayesev_modeling}
\begin{split}
p(\pmb{f}|z,M) &= \int p(\pmb{f}, \pmb{\alpha}|z,M) d\pmb{\alpha}\\ 
&= \int  p(\pmb{f}|\pmb{\alpha}, z,M) p(\pmb{\alpha} | z,M) d\pmb{\alpha}    
\end{split}
\end{equation} 
which is commonly referred as the Bayes evidence. We assume the likelihood $p(\pmb{f}|\pmb{\alpha}, z,M)$ is given by a normal multivariate distribution,
\begin{equation}\label{bayesevlikelihood}
\begin{split}
  p(\pmb{f}|\pmb{\alpha}, z,M) =& \frac{1}{\sqrt{(2\pi)^{d}}\prod_k \sigma(f_k)} \times \\
  &\exp\left[-\frac{1}{2}\sum_{k} \left(\frac{f_k}{\sigma(f_k)} - \sum_{j=1}^{n} \alpha_j\,\frac{t_{jk}}{\sigma(f_k)} \right)^2\right]   
\end{split}
\end{equation}
where $k$ runs over the bands, $j$ runs over the templates in the model $M$, $t_{jk}$ is the flux for band $k$ and template $t_j$ in the model, and $\sigma(f_k)$ is the measured flux error for band $k$. 

The integral in Equation~\ref{bayesev_modeling} requires that we specify a prior on the model parameters $p(\pmb{\alpha} | z,M)$. We distinguish between continuum and emission line SEDs in the prior,
\begin{equation} \label{prior_templates}
\begin{split}
     p(\pmb{\alpha} | z,M) =& p(\alpha_{0}^{\mathrm{Cont}}| z,M)\times \\
     &p(\alpha_{1}^{\mathrm{Cont}}| z,M)\times \\
     &p(\alpha^{\mathrm{EL}} |\alpha_{0}^{\mathrm{Cont}},\alpha_{1}^{\mathrm{Cont}}, z,M)  
\end{split}
\end{equation}
where we choose $p(\alpha_{0}^{\mathrm{Cont}}, z,M)$ and $p(\alpha_{1}^{\mathrm{Cont}}, z,M)$ to be top hat functions,
\begin{equation}\label{continuum_alphas_prior}
\begin{split}
    p(\alpha_{0}^{\mathrm{Cont}}| z,M) &\propto \Theta\left(\alpha_{0}^{\mathrm{Cont}}\right) \Theta\left(\Delta_0 - \alpha_{0}^{\mathrm{Cont}}\right) \\
    p(\alpha_{1}^{\mathrm{Cont}}| z,M) &\propto \Theta\left(\alpha_{1}^{\mathrm{Cont}}\right) \Theta\left(\Delta_1 - \alpha_{1}^{\mathrm{Cont}}\right)
\end{split}
\end{equation}
with $\Theta$ the Heaviside step function. Ideally, the prior for the continuum amplitudes (Equation~\ref{continuum_alphas_prior}) should contain information about the luminosity function. We use a top hat prior between 0 and a maximum flux of $(\Delta_0,\Delta_1)$, which regards values outside these bounds as unphysical. The values $(\Delta_0,\Delta_1)$ are set such that the maximum flux does not exceed a given maximum luminosity threshold in the Subaru $i$-band, and the thresholds are calculated from the Subaru i-band absolute magnitude which is derived from the data and the best model (see Section~\ref{sec:prior_template_params}).


The prior probability term $p(\alpha^{\mathrm{EL}} |\alpha_{0}^{\mathrm{Cont}},\alpha_{1}^{\mathrm{Cont}}, z,M)$ is one of the key ingredients of our modeling, which constrains the colours that arise from emission lines with respect to continuum flux. We estimate this prior using data, modeling a relation between the luminosity in the ultraviolet and the luminosity of the OII line, which we describe in detail in Section~\ref{sec:prior_template_params}.

We note that the Bayesian integral of Equation~\ref{bayesev_modeling} is a generalization of previous work \citep[see \textsc{EAZY} or \textsc{BCNZ2} codes][]{eazy,PAUSphotoz}, which approximate the integral with the maximum likelihood 
\begin{equation} \label{chi2likelihood}
p(\pmb{f}|z,M) \approx p(\pmb{f}|\pmb{\alpha}_{\mathrm{max}}, z,M)
\end{equation}
where $\pmb{\alpha}_{\mathrm{max}}$ are the maximum likelihood values of the parameters within the positive orthant. We will use this approximation only when calibrating the prior for the emission lines and in the zero point calibration step (see Section~\ref{sec:prior_template_params}\&\ref{sec:zeropoint_modeling}). To find the maximum likelihood values $\pmb{\alpha}_{\mathrm{max}}$ we will use the minimization algorithm from \textsc{bcnz2} \citep{PAUSphotoz}. We reproduce this algorithm in Appendix~\ref{sec:algorithm_minimization}.

To efficiently compute the integral from Equation~\ref{bayesev_modeling} we have implemented a code based on a Gaussian integral algorithm from \cite{Genz}. 
Details of the algorithm are explained in Appendix~\ref{sec:algorithm_integral}\&\ref{sec:genz_algorithm}.

\subsection{Galaxy SED} \label{sec:galaxysed_modeling}

\subsubsection{Continuum templates}

\begin{table}
\begin{tabular}{ll}
\hline Groups of Continuum templates & Extinction Laws   \\
 \hline  
0) BC$(0.008,10)$,  BC$(0.008,13)$ & None \\
1) BC$(0.008,8)$,  BC$(0.008,10)$ & None \\
2) BC$(0.008,6)$,  BC$(0.008,8)$ & None \\
3) BC$(0.008,4.25)$,  BC$(0.008,6)$ & None \\
4) BC$(0.008,2.6)$,  BC$(0.008,4.25)$ & None \\
5) BC$(0.02,10)$,  BC$(0.02,13)$ & None \\
6) BC$(0.02,8)$,  BC$(0.02,10)$ & None \\
7) BC$(0.02,6)$,  BC$(0.02,8)$ & None \\
8) BC$(0.02,4.25)$,  BC$(0.02,6)$ & None \\
9) BC$(0.02,2.6)$,  BC$(0.02,4.25)$ & None \\
10) Ell1,  Ell4 & None \\
11) Ell4,  Ell7 & None \\
12) Ell7,  Sc  & None \\
13) Sc,  SB0 & None, Prevot \\
14) SB0,  SB4  & None, Prevot \\
 \multirow{2}{*}{15) SB4,  SB8} & None, Calzetti, Calz.+Bump1,\\
 ~ &  Calz.+Bump2 \\
 \multirow{2}{*}{16) SB8,  SB11} & None, Calzetti, Calz.+Bump1,\\
 ~ &  Calz.+Bump2 \\
\end{tabular}
\caption{List of continuum synthetic templates considered in different models $M$. We use continuum templates used in \citet{COSMOS2015}. The elliptical (Ell) and spiral (Sc) templates were generated by \citet{Polleta2007}, while the starburst models (SB) were generated by \citet{BC2003}. The additional BC03 templates (BC) have their metallicity ($Z$) and age (Gyr) specified in parentheses, and were introduced in \citet{Ilbert13}. We apply reddening to the continuum SEDs using extinction laws for spiral and starburst templates, using a grid with 10 different $E(B-V)$ values spaced by 0.05 from 0.05 to 0.5 (see text for further details). In total, there are 97 combinations of continuum templates with different extinction laws and extinction values.
}
\label{tab:bayesev_models}
\end{table}

Here we describe the continuum galaxy SED templates used in this work. We will use a library of synthetic SED templates generated using recipes from \cite{BC2003} and \cite{Polleta2007}. This library, which is similar to the ones used in \cite{lephare,COSMOSILBERT,Ilbert13,COSMOS2015,PAUSphotoz}, contains a set of Elliptical, Spiral and Starburst synthetic templates. The templates in this library do not account for the light attenuation due to the internal dust present in each galaxy. Generally, this effect varies in each galaxy. Following the aforementioned references we model extinction using an extinction law $k(\lambda)$ and a color excess $E(B-V)$ that adds this effect to the template as
\begin{equation} \label{extinction}
F_{\textrm{observed}} (\lambda) = F_{\text{no dust}} (\lambda) \times 10^{-0.4 \, E(B-V) \, k(\lambda)}.
\end{equation}
We include this effect by modifying our default templates and generating new ones with different amounts of dust attenuation, since we cannot parametrize it linearly. For the starburst galaxies we will use the extinction law from \cite{Calzetti_extinction}, while for spirals we will model dust attenuation using the \cite{Prevot_extinction} law. We will not add extinction to the reddest galaxy templates, like ellipticals templates. We include two modifications of the Calzetti law with an additional bump around $2175\angstrom$ \citep[see section 3.4 of][]{COSMOSILBERT} which was not prominent in the original calibration of the law, but which has later been found in starburst galaxies \citep{Stecher65,2011ApJ...733...91X,1986ApJ...307..286F,2007ApJ...663..320F}. We use two different amplitudes of the bump following \cite{COSMOS2015,PAUSphotoz}. We generate a grid of templates with different $E(B-V)$ values ranging from 0.05 to 0.5 in steps of 0.05. We implement the IGM absorption using the analytical correction from \cite{Madau95}.

As mentioned in the previous section, each model $M$ used in this work contains two continuum templates. The complete list of groups of two continuum templates is shown in Table~\ref{tab:bayesev_models}. Some of the groups exist with different extinction laws created with a range of different $E(B-V)$ values, as indicated in the table. We have tested some combinations of three continuum templates but found it to have little impact on the probability. The Bayes evidence has preference for simpler models \citep[an effect often named as Bayesian Occam's Razor, e.g.][]{Ghahramani12} so that models with more templates than needed to describe the data naturally get a lower probability. A group of at least two continuum templates guarantees a more continuous coverage in color space. We leave an exploration of other combinations of templates, and the addition of different synthetic templates to future work.

\begin{table}\centering
\begin{tabular}{ccc}
\hline  \vspace{0.1cm} Line &   $\lambda[\angstrom]$   \\
 \hline   Ly$_{\alpha}$ & 1215.7   \\
 OII & 3726.8  \\
OIII$_1$ & 4959\\
OIII$_2$ &  5007  \\
H$_{\alpha}$ &  6562.8  \\
H$_{\beta}$ &  4861 \\
NII$_1 $&  6548  \\
NII$_2$ &  6583  \\
SII$_1$ &  6716.4 \\
SII$_2$ &  6730.8 
\end{tabular}
\caption{List of emission lines included in the SED modeling. The lines are modeled with a Gaussian distribution with a width of $10\angstrom$ centered around the air wavelengths shown in the second column.}
\label{emission_line_table}
\end{table}

\subsubsection{Emission lines} \label{sec:emission_line_modeling}

All the models $M$ include a third template which models the flux from emission lines. Table~\ref{emission_line_table} shows a list of the lines we include and the wavelength on which they are centered. We model each line with a Gaussian distribution of $10\angstrom$ width \citep[similar to \textsc{lephare},][which accounts for some Doppler broadening due to the galaxy's rotational velocity]{lephare}. Therefore, our model will integrate over different combinations of continuum flux and emission line flux, with a greater ability to describe the observed flux of each object than if we fixed these quantities.

We build the emission line template $t_{\mathrm{EL}}$ as
\begin{equation} \label{beta_definition}
    t_{\mathrm{EL}} \equiv \psi^{\OII} + \beta^{\OIII} \psi^{\OIII} + \beta^{\Halpha} \psi^{\Halpha} + \beta^{\Hbeta} \psi^{\Hbeta}
\end{equation}
where 
\begin{equation} \label{beta_ratios}
\begin{split}
    \psi^{\OII} &\equiv \OII + 2\Lya; \\
    \psi^{\OIII} &\equiv \frac{1}{3}\OIII_1 + \OIII_2; \\
    \psi^{\Halpha} &\equiv \Halpha + 0.35(\frac{1}{3}\mathrm{NII}_1 + \mathrm{NII}_2 + \mathrm{SII}_1 + \mathrm{SII}_2); \\
    \psi^{\Hbeta} &\equiv \Hbeta,
\end{split}
\end{equation}
The notation OII, for example, in the above equation means a Gaussian distribution centred at $\lambda=3726.8\angstrom$ and $10\angstrom$ width whose flux integrates to $10^{-17}\mathrm{erg}\,\mathrm{s}^{-1}\,\mathrm{cm}^{-2}$, and similar for the other lines using the wavelength values from Table~\ref{emission_line_table}. The parameters $\beta^{\OIII}$, $\beta^{\Halpha}$ and $\beta^{\Hbeta}$ indicate the relative amount of flux between ($\psi^{\OIII}$, $\psi^{\Halpha}$,  $\psi^{\Hbeta}$) and $\psi^{\OII}$ and are determined from data (see section~\ref{sec:prior_template_params}). 

The term $\psi^{\OII}$ contains the OII and $\Lya$ lines. We include the $\Lya$ line following previous photo-z analysis in the COSMOS field \citep{COSMOSILBERT,COSMOS2015} which used a fixed ratio of 2 between $\Lya$ and OII. The $\Lya$ line has a small impact since it only changes the flux of the Galex NUV filter for most of the galaxies in this analysis, which has the smallest signal to noise, and it only enters the CFHT u band above redshift 2. The term $\psi^{\OIII}$ contains the OIII doublet, where the factor $\frac{1}{3}$ comes from atomic physics \citep[e.g.][]{Storey00}. The term $\psi^{\Halpha}$ contains the $\Halpha$ line and the NII and SII doublets, where the factor $\frac{1}{3}$ also comes from atomic physics \citep{Storey00}. Note that the $\Halpha$ and NII lines are essentially blended together in our filter set, and we assign a fixed ratio of 0.35 to the NII lines with respect to $\Halpha$, although this ratio is smaller for lower stellar mass galaxies \citep[e.g.][]{2018ApJ...855..132F}. $\Halpha$ and SII doublet are distinguishable in different narrow bands until $z\sim0.3$, where they get redshifted outside of the PAUS narrow band coverage and become blended, so we model them together to have an homogeneous modeling at all redshifts. The $\Hbeta$ line is modeled separately in $\psi^{\Hbeta}$. The emission line modeling presented here can be improved further using known relations: the BPT diagram \citep{BPT} which establishes relations between the OIII, NII, $\Halpha$ and $\Hbeta$ lines; the intrinsic case B recombination Balmer decrement $(\Halpha/\Hbeta)=2.86$ \citep{Storey95,Moustakas2006}; or modeling SII and $\Halpha$ differently below and above $z\sim0.3$. We defer a thorough exploration of these possibilities for future work. 

Finally, we apply the same extinction law for the continuum templates and the emission line template, and modify the ratio values according to the attenuation from extinction. The amount of dust attenuation for emission lines in nebular regions can be different than that from stellar continuum \citep[see][]{Calzetti94,Calzetti_extinction,Puglisi16,Saito20}. Measuring the ratios $\beta^{\OIII}$, $\beta^{\Halpha}$, $\beta^{\Hbeta}$ and the prior of the emission line template $p(\alpha^{\mathrm{EL}} |\alpha_{0}^{\mathrm{Cont}},\alpha_{1}^{\mathrm{Cont}}, z,M)$ from data assuming the same amount of dust attenuation in the model yields a more consistent flux model overall (see section~\ref{sec:prior_template_params}).

\subsection{Prior on  template parameters} \label{sec:prior_template_params}

In this section we describe how we estimate several parameters in the model using spectroscopic data. To recap, we need to estimate the parameters  $(\Delta_0,\Delta_1)$ introduced in the continuum templates prior (Equation~\ref{continuum_alphas_prior}), describe the prior on the emission line template $p(\alpha^{\mathrm{EL}} |\alpha_{0}^{\mathrm{Cont}},\alpha_{1}^{\mathrm{Cont}}, z,M)$ (Equation~\ref{prior_templates}) and set the parameters ($\beta^{\OIII}$, $\beta^{\Halpha}$, $\beta^{\Hbeta}$) from Equation~\ref{beta_ratios}.

For each galaxy with a confident\footnote{For this exercise we use only galaxies in the zCOSMOS-Bright DR3 release with a confidence flag $c$ in the set: [3.x, 4.x, 2.4, 2.5, 1.5, 9.3, 9.4, 9.5, 18.3, 18.5]} spectroscopic redshift in the zCOSMOS-Bright spectroscopic catalog (see section~\ref{sec:spectroscopic_data}), we carry out the following steps
\begin{enumerate}
    \item Find the model $M$ with the largest Bayes evidence at the galaxy's spectroscopic redshift (Equation~\ref{bayesev_modeling}).
    \item For the most probable model $M^{\mathrm{max}}$, we find the maximum likelihood parameters $\pmb{\alpha}_{\mathrm{max}}$ using the minimization algorithm (Equation~\ref{chi2likelihood}).
    \item Subtract from the data the estimated continuum using the continuum $\pmb{\alpha}_{\mathrm{max}}$ parameters.
    \item Produce Gaussian realizations centered at the subtracted flux using the measured flux error. For each realization, find the best fit values for that galaxy: ($\alpha_{g}^{\mathrm{EL}}$, $\beta_{g}^{\OIII}$, $\beta_{g}^{\Halpha}$, $\beta_{g}^{\Hbeta}$). Note that since we separately model additional internal dust reddening, these values are extinction free by definition.
    \item Estimate the mean and standard deviation of the parameters ($\alpha_{g}^{\mathrm{EL}}$, $\beta_{g}^{\OIII}$, $\beta_{g}^{\Halpha}$, $\beta_{g}^{\Hbeta}$) from the previous step.
\end{enumerate}
From step (ii) we can also obtain absolute magnitudes for every galaxy, which we can correct for internal dust extinction using the best model (and the best extinction parameters). Of interest for us are the absolute magnitudes $\MNUV$, $M_{\mathrm{I}}$, which stand for the GALEX NUV and Subaru $i$ bands. In summary, this algorithm provides the parameters ($\langle\alpha_{g}^{\mathrm{EL}}\rangle$, $\langle\beta_{g}^{\OIII}\rangle$, $\langle\beta_{g}^{\Halpha}\rangle$, $\langle\beta_{g}^{\Hbeta}\rangle$,$\MNUV$, $M_{\mathrm{I}}$) for each of the aforementioned galaxies.

The distribution of $M_I$ peaks around magnitude -22 in our sample, and we find no galaxies brighter than -26. We assume this value to be the brightest magnitude a galaxy could be in this band, and set the upper limit of the top hat prior of the continuum templates $(\Delta_0,\Delta_1)$ (Equation~\ref{continuum_alphas_prior}) to fulfil this condition at all redshifts.

\begin{figure}
	\includegraphics[width=\columnwidth]{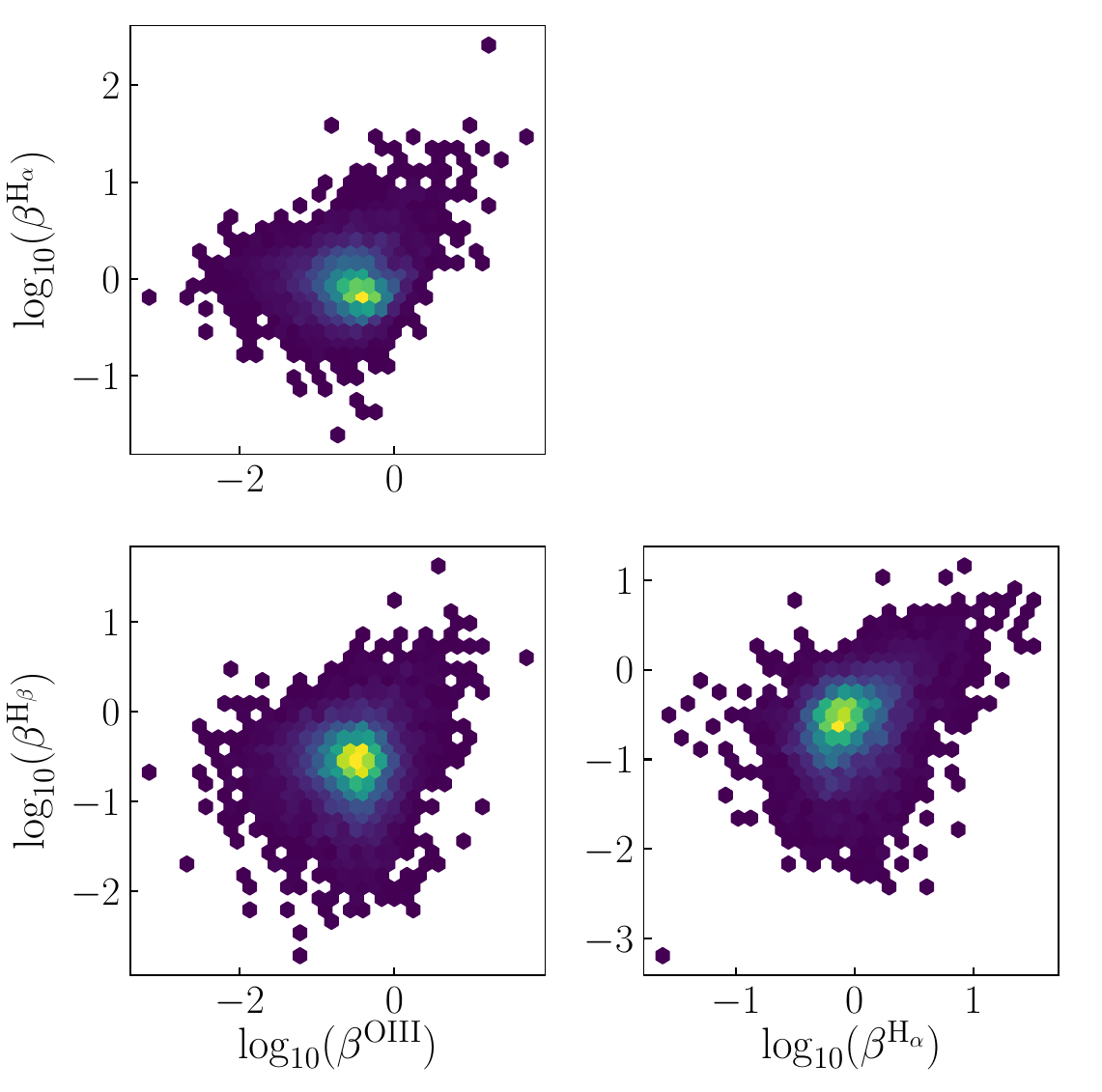}
    \caption{This figure shows the line flux ratio between (mainly) the $\OIII$, $\Halpha$ and $\Hbeta$ lines with respect to the $\OII$ line, as the density plot of the logarithmic measured values of $\beta^{\OIII}$, $\beta^{\Halpha}$ and $\beta^{\Hbeta}$ (Equation~\ref{beta_definition}) which give the relative amount of flux between different emission lines, as defined by Equation~\ref{beta_ratios}. The combined photometry of COSMOS and PAUS is used to make the measurement for each object. We fit models of the emission lines to a continuum subtracted measured flux for a subset of objects with spectroscopic redshift, correcting for internal dust extinction (see section~\ref{sec:prior_template_params}).}
    \label{fig:beta_ratios}
\end{figure}

Figure~\ref{fig:beta_ratios} shows a density plot of the estimated mean values $\langle\beta_{g}^{\OIII}\rangle$, $\langle\beta_{g}^{\Halpha}\rangle$, $\langle\beta_{g}^{\Hbeta}\rangle$. For many of the galaxies, the measured signal-to-noise ratio on these parameters is low. Therefore, the density plot is a convolution of some underlying distribution convolved with this noise. We do not attempt to recover the true distribution in this work, but instead just measure the median and $\sigma_{68}$ of the marginal noisy distribution, finding:
\begin{equation} \label{estimated_beta_parameters}
\begin{split}
    \log_{10}(\beta_{g}^{\OIII}) &= -0.50\pm0.35 \\
    \log_{10}(\beta_{g}^{\Hbeta}) &= -0.56\pm0.34 \\
    \log_{10}(\beta_{g}^{\Halpha}) &= -0.08\pm 0.24 \\
\end{split}
\end{equation}
For each pair of continuum SEDs we add one emission line model with parameters $\beta^{\OIII}$, $\beta^{\Halpha}$, $\beta^{\Hbeta}$ equal to the median values from Equation~\ref{estimated_beta_parameters}. In order to account for the breadth of the distribution we include 6 additional models per continuum group, each with one of the $\beta$ parameters set at a value two times the $\sigma_{68}$ with respect to its median\footnote{These six additional models can be expressed as ([2,0,0],[0,0,2],[0,2,0],[-2,0,0],[0,0,-2],[0,-2,0]), where for example [2,0,0] would mean ($\beta_{g}^{\Hbeta}$, $\beta_{g}^{\Halpha}$) are set at their median value, while $\beta_{g}^{\OIII}$ is set at at its median value plus two times the measured $\sigma_{68}$.}. In total, we have 679 different models $M$, each with different continuum or emission line templates.

It is worth noting that step (i) in this section requires an initial assumption about the values of ($\beta^{\OIII}$, $\beta^{\Halpha}$, $\beta^{\Hbeta}$) and the prior $p(\alpha^{\mathrm{EL}} |\alpha_{0}^{\mathrm{Cont}},\alpha_{1}^{\mathrm{Cont}}, z,M)$. We initially fix these values to the line flux ratio values from \cite{COSMOSILBERT}. After the initial run, we repeat the process to measure all of these parameters and calibrate the prior (which is described later in this section). We repeat this process a couple of times, after which we find the values do not change. All the values reported in this section are the final values, which are used in the remainder of this work.

\subsubsection{Prior $p(\alpha^{\mathrm{EL}} |\alpha_{0}^{\mathrm{Cont}},\alpha_{1}^{\mathrm{Cont}}, z,M)$}

Our model requires a prior on the emission line template free amplitude $\alpha^{\mathrm{EL}}$. We will use the known correlation between the UV luminosity of a galaxy and its OII emission line flux, which has been used before in photometric redshift estimation \citep{Kennicutt1998, COSMOSILBERT}, to build a model between $\alpha^{\mathrm{EL}}$ and the absolute magnitude $\MNUV$. We assume the following relation, 
\begin{equation}\label{priorEL}
\begin{split}
    p(\eta|\MNUV)  \sim& \mathcal{N}(\mu=a \MNUV + b, \sigma=c), \\
    \eta \equiv& -2.5\log_{10}(\alpha^{\mathrm{EL}})-\mathrm{DM},
\end{split}
\end{equation}
where DM is the distance modulus, $\mathcal{N}$ is a Gaussian distribution, and $a,b,c$ are parameters to be determined from data. The parameter $\eta$ can be interpreted as an emission line absolute magnitude. Therefore, we assume there is a linear relation between $\eta$ and $\MNUV$ with an intrinsic scatter perpendicular to the correlation given by a normal distribution. In the model, the absolute magnitude $\MNUV$ is a function of the three free amplitudes, $\MNUV=\MNUV(\alpha_{0}^{\mathrm{Cont}},\alpha_{1}^{\mathrm{Cont}}, \alpha^{\mathrm{EL}})$. Therefore Equation~\ref{priorEL} describes the probability of $\alpha^{\mathrm{EL}}$ given $(\alpha_{0}^{\mathrm{Cont}},\alpha_{1}^{\mathrm{Cont}},z)$ which is our probability $p(\alpha^{\mathrm{EL}} |\alpha_{0}^{\mathrm{Cont}},\alpha_{1}^{\mathrm{Cont}}, z,M)$ for a given model $M$. 

\begin{figure}
	\includegraphics[width=\columnwidth]{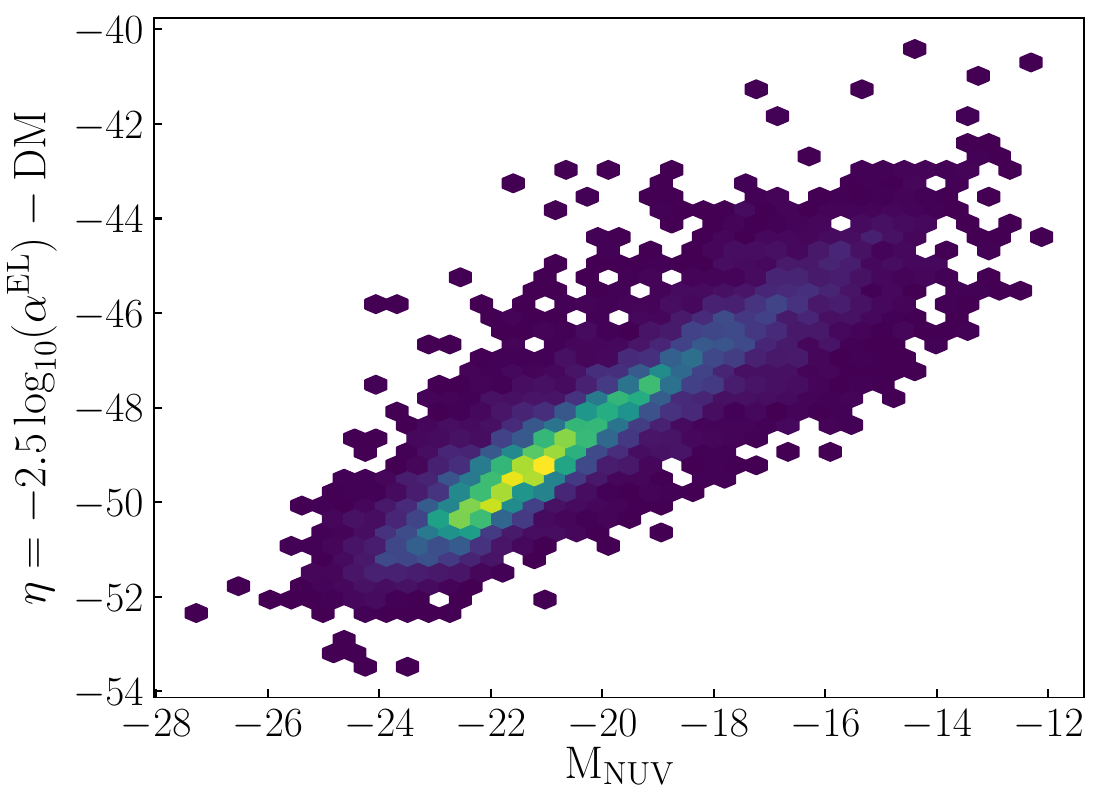}
    \caption{This figure shows the correlation between the OII line flux and the UV light as a density plot of the measured values of $\eta\equiv -2.5\log_{10}(\alpha^{\mathrm{EL}})-\mathrm{DM}$ and $\MNUV$ (Equation~\ref{priorEL}) for a subset of objects with spectroscopic redshift, where DM is the distance modulus. We expect these variables to be correlated since the $\OII$ line flux is correlated with the ultraviolet luminosity of a galaxy. We model this as a linear relation with an intrinsic Gaussian scatter, and find the most likely parameters of this model to calibrate the prior between continuum and emission line templates, which is one of the key ingredients in our redshift estimation model (see section~\ref{sec:prior_template_params}). The measurements have been corrected for internal extinction.}
    \label{fig:alpha_MUV_relation}
\end{figure}

Figure~\ref{fig:alpha_MUV_relation} shows the relation between $\eta$ and $\MNUV$ as a density plot of the measured values for each galaxy, with $\alpha^{\mathrm{EL}}$ expressed in units of $10^{-17}\mathrm{erg}\,\mathrm{s}^{-1}\,\mathrm{cm}^{-2}$. Similar to Figure~\ref{fig:beta_ratios}, the plot does not show the relation directly, since it is convolved with the measurement noise of both variables. The inference of the parameters $(a,b,c)$ in Equation~\ref{priorEL} has to be done carefully to avoid introducing a bias since the data is noisy in both axes \citep[for more details see][]{Kelly07,Hogg10fitaline}. 

We will infer the parameters $\theta\equiv(a,b,c)$ by writing the likelihood of the observations given the parameters, following a model similar to \cite{Kelly07}. Let $\eta_g$, $\mathrm{M}_g$ be the true values of the variables $\eta$, $\MNUV$ for galaxy $g$, and $\hat{\eta}_g$, $\hat{\mathrm{M}}_g$ be their noisy observational counterparts. We will assume $\mathrm{M}_g$ follows from a probability distribution $p(\mathrm{M}_g|\xi)$, where $\xi$ are the parameters of the distribution. The joint distribution of $\eta_g$ and $\mathrm{M}_g$ is then $p(\eta_g,\mathrm{M}_g|\theta,\xi) = p(\eta_g|\mathrm{M}_g,\theta)p(\mathrm{M}_g|\xi)$. We will assume a Gaussian and independent measurement error in $\hat{\eta}_g$ and $\hat{\mathrm{M}}_g$, so that $p(\hat{\eta}_g,\hat{\mathrm{M}}_g|\eta_g,\mathrm{M}_g) = p(\hat{\eta}_g|\eta_g)p(\hat{\mathrm{M}}_g|\mathrm{M}_g)$ are two Gaussian distributions with means $(\eta_g,\mathrm{M}_g)$ and variances $(\sigma^2(\hat{\eta}), \sigma^2(\hat{\mathrm{M}}_g))$. Therefore, we can hierarchically express the model as
\begin{equation}
\begin{split}
    \mathrm{M}_g &\sim p(\mathrm{M}_g|\xi) \\
    \eta_g|\mathrm{M}_g &\sim \mathcal{N}(a \mathrm{M}_g + b, c) \\
    \hat{\eta}_g,\hat{\mathrm{M}}_g|\eta_g,\mathrm{M}_g &\sim \mathcal{N}(\eta_g, \sigma(\hat{\eta}))\times\mathcal{N}(\mathrm{M}_g, \sigma(\hat{\mathrm{M}}_g)))
\end{split}
\end{equation}
The likelihood function of the measured data $p(\hat{\eta}_g,\hat{\mathrm{M}}_g|\theta,\xi)$ can be obtained by integrating the complete data likelihood over the missing data $\eta_g,\mathrm{M}_g$
\begin{equation}
\begin{split}
    p(\hat{\eta}_g,\hat{\mathrm{M}}_g|\theta,\xi) =& \int\int p(\hat{\eta}_g,\hat{\mathrm{M}}_g,\eta_g,\mathrm{M}_g|\theta,\xi)\,d\eta_g d\mathrm{M}_g  \\ 
    =& \int\int p(\hat{\eta}_g,\hat{\mathrm{M}}_g|\eta_g,\mathrm{M}_g)p(\eta_g|\mathrm{M}_g,\theta) \\
    &\times p(\mathrm{M}_g|\xi) \,d\eta_g d\mathrm{M}_g.
\end{split}
\end{equation}
In this work we will model $p(\mathrm{M}_g|\xi)$ with a mixture of $K$ Gaussian distributions,
\begin{equation}
    p(\mathrm{M}_g|\xi) = \sum_{k=1}^{K} \frac{\pi_k}{\sqrt{2\pi\tau_k^2}} \exp\left\{-\frac{1}{2}\frac{(\mathrm{M}_g-\mu_k)^2}{\tau_k^2} \right\},
\end{equation}
where $\sum_k \pi_k = 1$. Defining $\pi\equiv(\pi_1,\ldots,\pi_K)$, $\mu\equiv(\mu_1,\ldots,\mu_K)$ and $\tau\equiv(\tau_1,\ldots,\tau_K)$, note that we have $\xi=(\pi, \mu, \tau)$. This mixture model is flexible enough to describe a wide variety of distributions, and it is also convenient since it simplifies the mathematics for writing the likelihood of the measured data \citep[see][]{Kelly07}. Assuming the data for different galaxies is statistically independent, the full data likelihood is the product of the measurement likelihood of each galaxy
\begin{equation} \label{complete_likelihood_data}
\begin{split}
    p(\{\hat{\eta}_g\},\{\hat{\mathrm{M}}_g\}|\theta,\xi) =& \prod_{g=1}^{n}\sum_{k=1}^{K} \frac{\pi_k}{2\pi|V_{g,k}|^{1/2}} \\
    &\times \exp\left\{-\frac{1}{2}(\pmb{z}_g-\zeta_k)V^{-1}_{g,k}(\pmb{z}_g-\zeta_k)^{\intercal} \right\} ,
\end{split}
\end{equation}
with
\begin{equation}
\begin{split}
    z_g &= (\hat{\mathrm{M}}_g,\hat{\eta}_g), \\ 
    \zeta_k &= (a\mu_k + b, \mu_k), \\
    V_{g,k} &=\begin{pmatrix} a^2\tau^2_k+c^2+\sigma^2(\hat{\eta}_g)&a\tau^2_k\\a\tau^2_k&\tau^2_k+\sigma^2(\hat{\mathrm{M}}_g)) \end{pmatrix} .
\end{split}
\end{equation}
We fix the number of mixture Gaussians to $K=2$ (although we have verified that the results for $\theta$ do not change if $K=\{2,3,4,5\}$). We maximize the likelihood in Equation~\ref{complete_likelihood_data} and find the most likely parameters $\theta^{\mathrm{ML}}$ to be
\begin{equation}\label{measured_UV_OII_params}
    a^{\mathrm{ML}} = 0.750 \\
    b^{\mathrm{ML}} = -33.38 \\
    c^{\mathrm{ML}} = 0.327 \\
\end{equation}
We will use these values for all the results in this work. Finally, we numerically integrate the prior, Equation~\ref{prior_templates}, to compute the prior normalisation, which is needed for the Bayes evidence, using a Metropolis-Hastings integration algorithm.

\subsection{Systematic zero points offsets} \label{sec:zeropoint_modeling}

A common approach in the literature is to find systematic relative (not global) zero-points between different bands before running the photo-z algorithm \citep[e.g.][]{BPZ,BPZCoe06,cfht_hilde,alhambra,COSMOS2015,PAUSphotoz}. This attempts to optimise the colours predicted by the model in comparison to the observed colours in the data to improve the photo-$z$ \citep[\textit{e.g.}][]{Dahlen2013}. A zero point offset does not need to come from the zero point estimation itself, but can also be due to an incorrect PSF modelling (\citealt{cfht_hilde}), and from incorrect or missing templates. 

We calibrate the systematic offsets with the same spectroscopic catalogue described in section~\ref{sec:prior_template_params}. We also use a similar algorithm. For every galaxy, we find the model $M$ with the largest Bayes evidence $M^{\mathrm{max}}$ and its most likely parameters $\pmb{\alpha}_{\mathrm{max}}$. We build the most likely fluxes $T_i$ according to $M^{\mathrm{max}}$ and $\pmb{\alpha}_{\mathrm{max}}$.

We assume the measured and predicted fluxes to be statistically independent for every galaxy and band, and find the offsets $\{\kappa_j\}$ that maximise the likelihood
\begin{equation} \label{zeropopint_kappa}
\begin{split}
     p(\{f_g\},\{\sigma_g\}, \{T_{g,k}\}|\{\kappa_j\}) \approx& \prod_k \prod_g p(f_g,\sigma_g, T_{g,k}|\kappa_j)   \\  
     =&\prod_k \prod_g \mathcal{N}(f_g\kappa_j- T_{g,k},\kappa_j\sigma_g).
\end{split}
\end{equation}
We apply the offsets (or factors for fluxes) to the data and run again, repeating the process until convergence. We exclude galaxies with a very bad fit ($\chi^2>120$, with $\sim 63$ degrees of freedom). The values of the offsets can be found in Appendix~\ref{app:sysoffs}, in Table~\ref{tab:offsets} and Figure~\ref{fig:offsets}. It is worth noting that the calibration of the prior described in section~\ref{sec:prior_template_params} and the offset calibration described in this section depend on each other. We hierarchically run each part of the calibration, using the prior parameters and the zero point offsets from the previous step. We perform this a couple of times.

\subsection{Population prior and redshift posterior} \label{sec:population_prior}

To obtain the redshift posterior $p(z|\pmb{f})$ for each galaxy we need to calculate Equation~\ref{redshift_posterior}, which requires us to know the population's distribution over different redshifts and models, $p(z,M)$. This quantity is unknown a priori, and previous template codes and analysis have made different assumptions, such as assuming it is uniform, or introducing analytical functions with hyperparameters calibrated with a spectroscopic population. Once a target population has been identified, the posterior on the redshift and model of each galaxy in the population can be jointly and hierarchically inferred along with the population's distribution over different redshifts and models \citep[e.g.][]{leistedt2016}. This can be further extended to include a dependence of galaxy density on the line of sight position due to galaxy clustering \citep[see][]{SB19,ASBG}. 

The redshift posterior of a galaxy is not unique since it depends on the population to which it belongs, or in other words different galaxy sample selections will yield different $p(z,M)$, and thus a different posterior $p(z|\pmb{f})$ for each galaxy. One proposed application of this redshift sample is to empirically calibrate the redshift distribution of galaxy samples from weak lensing surveys. Various techniques exist \citep{Wright19KidsSOM,SB19,Buchs2019,ASBG,SVsanchez20} which write a probability relation between the weak lensing galaxies and the galaxies from the calibration samples. The most correct output for such studies would be to produce and release the full likelihood $p(z,M|\pmb{f})$ for each galaxy, so that the redshift posterior and population distribution can be inferred correctly for any galaxy sample. This is impractical since the likelihood contains over a million values for each galaxy. Instead, we will compute and release a pseudo probability $\tilde{p}(z|\pmb{f})$ defined as
\begin{equation} \label{pseudoposterior}
    \tilde{p}(z|\pmb{f}) = \sum_{\{M\}} p(z,M|\pmb{f}) \approx \sum_{\{M\}} p(\pmb{f}|z,M) p(M)
\end{equation}
where we marginalize over the 679 models $M$, explicitly assuming a uniform prior in redshift $p(z|M)$. This probability can be interpreted as an effective likelihood of a unique \textit{pseudo} model, since it is a weighted likelihood over different models and has no explicit redshift prior. This effective \textit{likelihood} $\tilde{p}$ can be used to infer the redshift distribution of a given population, and it is a good approximation when all the $p(\pmb{f}|z,M)$ which contribute significantly are similar, and given that $p(M)$ is close to the real distribution of the target population.

We calibrate $p(M)$ using a subset of the broad band colours we have available in data and the same colours predicted by each model. The details are given in Appendix~\ref{app:popprior}.

\subsection{Comparison to previous models} \label{sec:previous_work}

The flux model developed here shares several elements with those implemented in \textsc{bcnz2} \citep[used in][Er19]{PAUSphotoz} and \textsc{lephare} \citep[used in][\cosmos]{COSMOS2015}, and it is worth highlighting some differences between them. Regarding the SED templates, the galaxy continuum templates and dust extinction laws used are the same in all models, but the emission line templates implementation differs. Er19 and \cosmos~create emission line templates using fixed line flux ratios between several lines and the OII line, as measured by different spectroscopic surveys, and originally collected in \cite{COSMOSILBERT}. Here, we use a few different values for the line flux ratios of line templates that contain the OII, $\Halpha$, $\Hbeta$ and OIII emission lines (Equations~\ref{beta_definition},\ref{beta_ratios}\&\ref{estimated_beta_parameters}), measured directly from the 66 photometric bands and the best model. In \cosmos~ the emission line template was combined with the continuum template at three fixed amplitudes with respect to the continuum template, which were given by the correlation between the UV luminosity and the OII line from \cite{Kennicutt1998}. Er19 left the amplitudes of the templates free (with a nonnegativity constraint), and allowed for setups where the OIII doublet was an extra template, separated from the other lines. Here, each model contains one emission line template, and its amplitude is marginalized with a prior that also accounts for the distribution between UV luminosity and the OII line, which we measure from the 66 photometric bands (Equation~\ref{measured_UV_OII_params}). 

Finally, in \cosmos~ one single template which combined continuum and emission line flux was fitted to the data, and the best fitting amplitude was used to infer the likelihood at each redshift; in Er19 the combination of several continuum templates (6 to 10) and several emission line templates (0 to 2) were maximized, and the best fitting combination was used for the redshift inference; here, we marginalize over the amplitudes of two continuum templates and one emission line template with priors and compute the Bayesian integral for the redshift inference. 

\begin{figure}
	\includegraphics[width=\columnwidth]{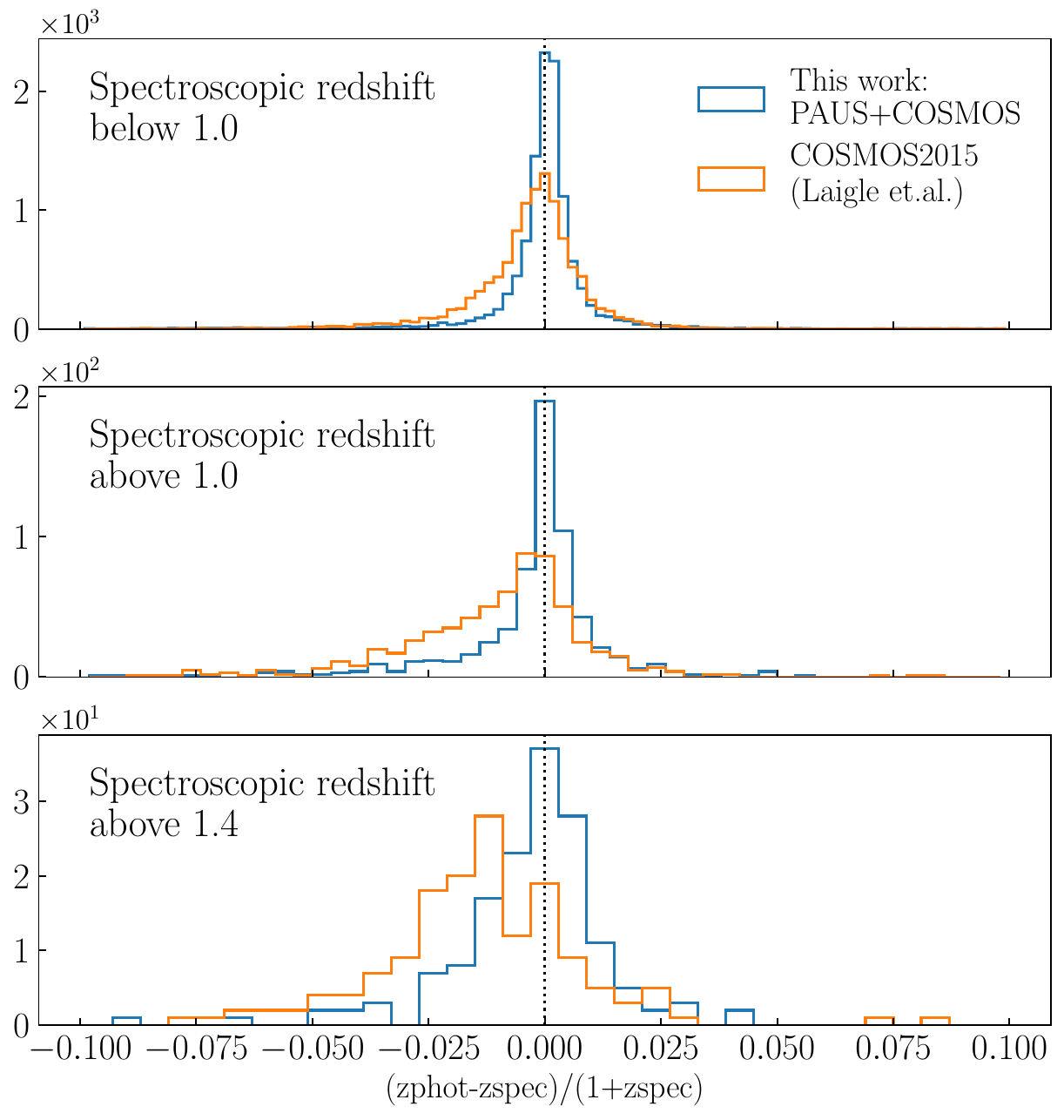}
    \caption{This figure shows the $\Delta_z\equiv
(z_{\mathrm{phot}}-z_{\mathrm{spec}})/(1+z_{\mathrm{spec}})$ histogram for spectroscopic objects below redshift 1 (top panel), for spectroscopic objects above redshift 1 (middle panel) and spectroscopic objects above redshift 1.4 (bottom panel). The photo-$z$ from this work are used in the orange histograms, while the blue histograms use the photo-$z$ from the public \cosmos~catalogue. The histograms clearly show the differences in the photo-$z$ precision, and also hint at a difference in the accuracy, \textit{i.e.} how well the distributions are centred on 0, which is shown by a vertical dashed line.}
    \label{fig:deltaz_zcuts}
\end{figure}

\begin{figure}
	\includegraphics[width=\columnwidth]{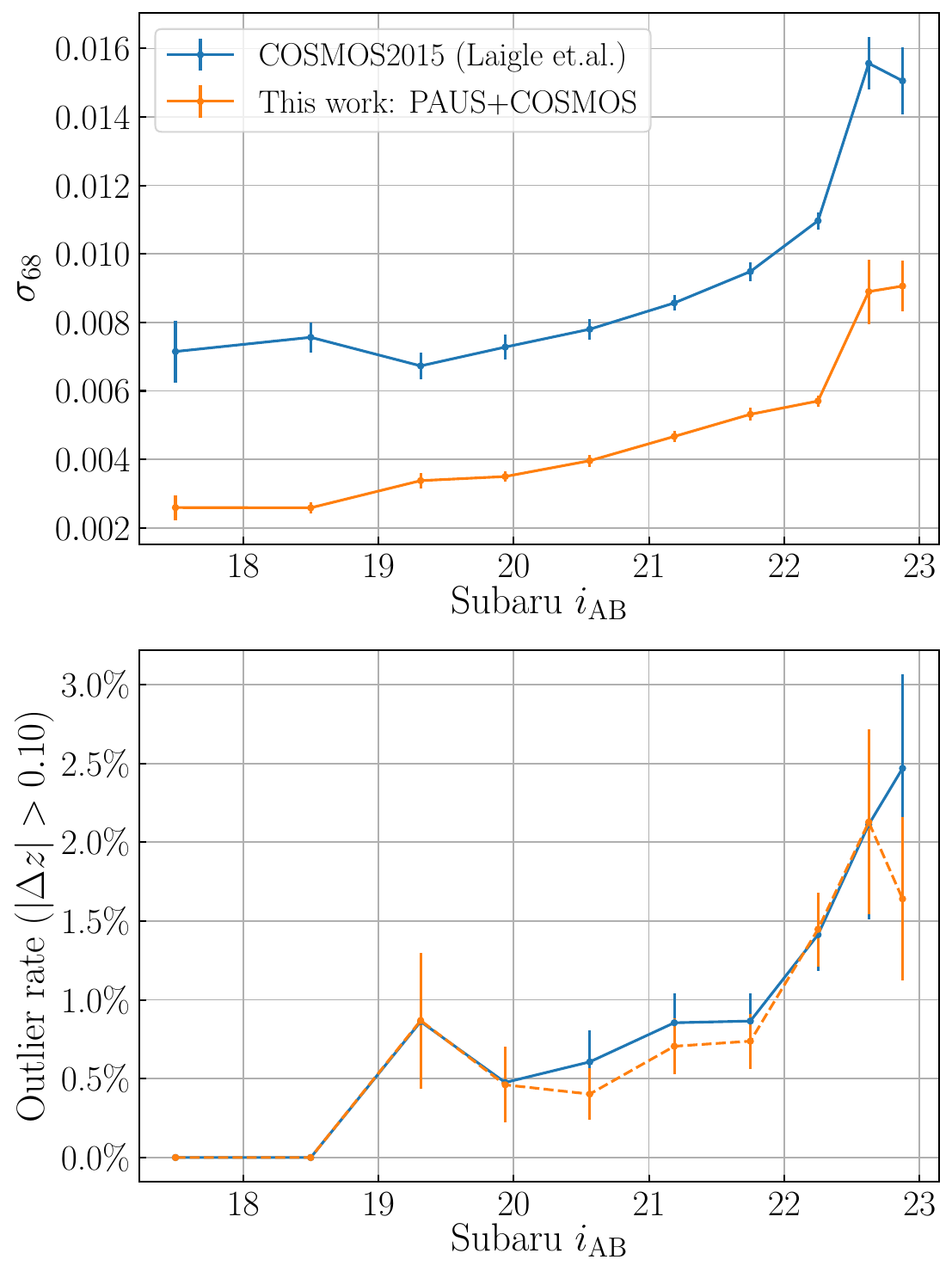}
    \caption{The precision of the photo-$z$ point estimates from this work (orange lines) with respect to the spectroscopic redshift catalog. The top panel shows the $\sigma_{68}$ (Eq.~\ref{sigma68}) of the $\Delta_z\equiv
(z_{\mathrm{phot}}-z_{\mathrm{spec}})/(1+z_{\mathrm{spec}})$ distribution as a function of $\iab$ magnitude, while the bottom panel shows the percentage of galaxies classified as photo-$z$ outliers, defined as objects that fulfill $|\Delta_z|>0.1$, also as a function of $\iab$ magnitude. We compute the same statistics using the photo-$z$ estimate from the \cosmos~public catalog (blue lines). We find a significant improvement in the redshift precision (lower $\sigma_{68}$) at all magnitudes considered in this work. The error bars are found by computing the dispersion of each metric when bootstrapping the objects in each magnitude bin.}
    \label{fig:sigma68OR}
\end{figure}

\section{Results}\label{sec:results}

In this section we present the photometric redshift measurements obtained using the model described in section~\ref{sec:methodology}, with the photometry from PAUS and COSMOS described in section~\ref{sec:data}. We will compare the redshift estimates with the spectroscopic catalog described in section~\ref{sec:spectroscopic_data}.

\subsection{Photometric redshift precision}

\begin{figure}
	\includegraphics[width=\columnwidth]{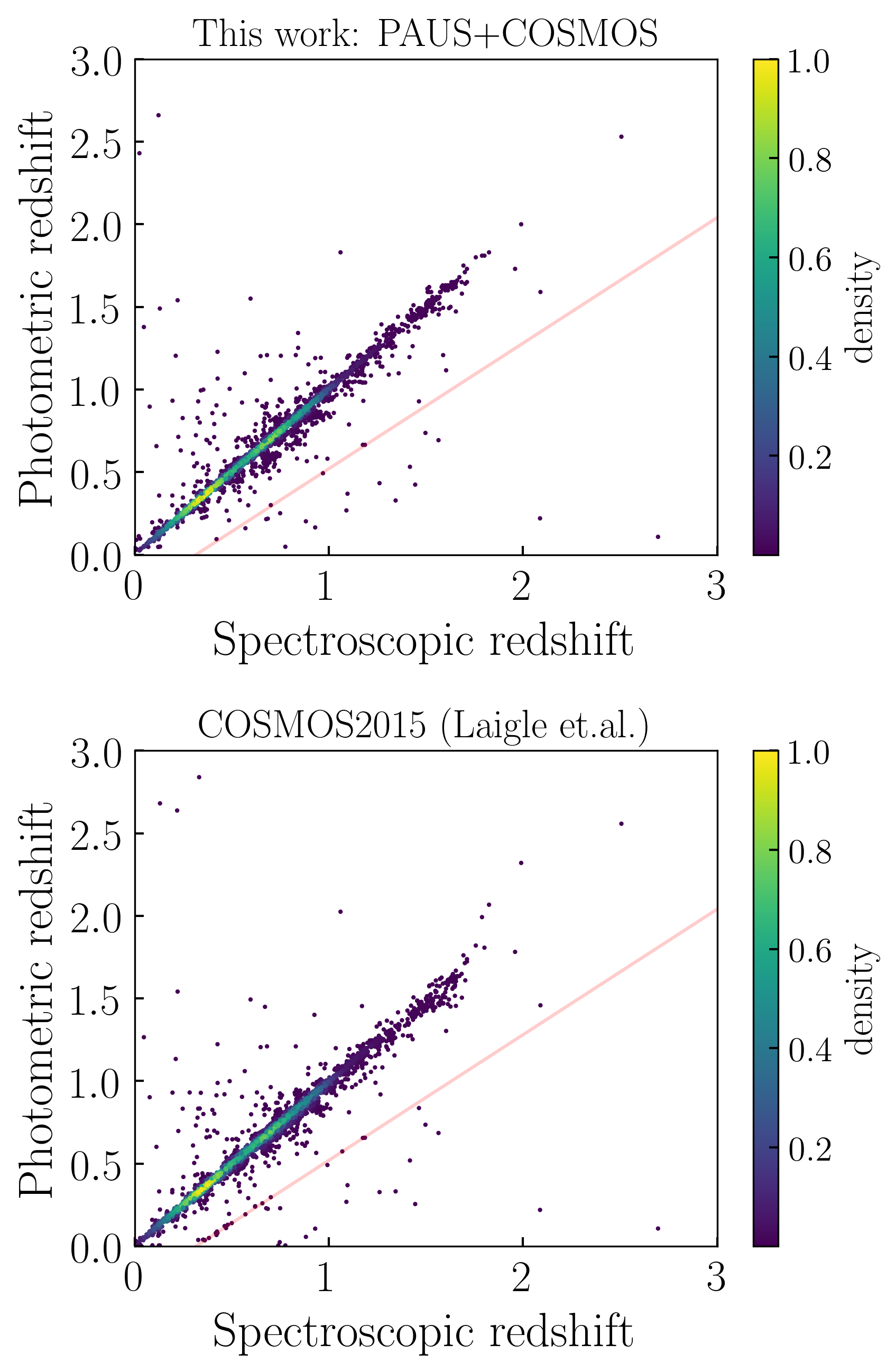}
    \caption{The scatter plot of the spectroscopic redshifts and the photo-$z$ point estimates of this work (top panel) and the photo-$z$ point estimates from \cosmos~(bottom panel). Points are colored according to the proximity (or density) of other nearby objects in this space. We find fewer outliers at $\Delta_z\approx-0.24$ (highlighted with a faint red line in both panels), which are consistent with a confusion between the $\OIII$ and $\Halpha$ lines, where $\Delta_z\equiv
(z_{\mathrm{phot}}-z_{\mathrm{spec}})/(1+z_{\mathrm{spec}})$.}
    \label{fig:scatter_redshift}
\end{figure}

We define our photo-$z$ point estimate $z_{\mathrm{phot}}$ as the mode of the redshift distribution of each galaxy $\tilde{p}(z|\pmb{f})$  (Equation~\ref{pseudoposterior}). When comparing to the \cosmos~photo-$z$ we will use the column \texttt{PHOTOZ} from their public catalog.

To assess the accuracy and precision of the photo-$z$ point estimates with respect to the spectroscopic point estimates, we consider the distribution of the following quantity: $\Delta_z\equiv
(z_{\mathrm{phot}}-z_{\mathrm{spec}})/(1+z_{\mathrm{spec}})$. We define two metrics to assess the photo-$z$ precision. One is the central dispersion of the $\Delta_z$ distribution, $\sigma_{68}$, defined as
\begin{equation} \label{sigma68}
\sigma_{68} \equiv \frac{P[84]-P[16]}{2}
\end{equation}
where $P[x]$ is the value of the distribution $\Delta_z$ for the percentile $x$, which is more robust to outliers than the standard deviation of the distribution. The second metric addressing the photo-$z$ precision is the outlier rate, which is the percentage of outlier galaxies, defined as galaxies that fulfil
\begin{equation}
\mathrm{Outlier}\equiv \frac{|z_{\mathrm{phot}}-z_{\mathrm{spec}}|}{1+z_{\mathrm{spec}}} > 0.1,
\end{equation}
which is similar in magnitude to what a galaxy would be considered as an outlier in typical lensing surveys \citep{Hildebrandt2017,Hoyle2017}. 

Figure~\ref{fig:deltaz_zcuts} visually shows the $\Delta_z$ distribution for the range $-0.1<\Delta_z<0.1$, highlighting that the $\Delta_z$ values from this new catalog are not only tighter, but also clearly less biased than \cosmos, specially for higher redshifts. Figure~\ref{fig:sigma68OR} shows the $\sigma_{68}$ (top panel) and outlier rate (bottom panel) for this work's photo-$z$ from the combination of PAUS and \cosmos~ photometry (orange lines) as a function of the Subaru $\iab$ (each point is a different magnitude bin). We find a $\sigma_{68}=0.0026$ at bright magnitudes of $\iab\sim18$ which increases up to $\sigma_{68}\sim0.009$ at $\iab\sim23$. There is a jump at $\iab\sim22.5$, $\sigma_{68}\sim0.0057$, where the spectroscopic sample completeness changes significantly since the zCOSMOS survey stops at $\iab=22.5$. 

The quasi spectroscopic precision at bright magnitudes is a common feature of analysis containing PAUS photometry \citep{PAUSphotoz,deepz}, since sharp features present in the galaxy spectrum can be precisely identified with the narrow bands, which have a FWHM of $100\angstrom$. For each magnitude bin, we compute the $\sigma_{68}$ using the photo-$z$ from the \cosmos~catalog (blue lines) for the same galaxies. We find a similar trend with magnitude, with $\sigma_{68}\sim0.007$ at the bright end, and $\sigma_{68}\sim0.015$ at the faint end. In comparison, we find this new catalog yields a redshift precision which is $1.66\times$ tighter than \cosmos~at $\iab\sim23$, and up to $3\times$ at bright magnitudes. We also find the same jump at $\iab=22.5$ with the \cosmos~photo-$z$, which means it is unrelated to the narrow band photometry from PAUS, and likely a feature of the spectroscopic sample. If the case, one explanation could be that the spectra is covering the galaxy population differently in the last two magnitude bins, since the completeness is lower (Figure~\ref{fig:spec_compl}), or the spectroscopic redshifts could be noisier and have a worse performance.

We find a very similar outlier rate for this analysis and \cosmos~(bottom panel), which is $\sim 1\%$ at $\iab\leq22.5$, and increases to $\sim2\%$ at $\iab\sim23$, with slightly smaller values for our work. Figure~\ref{fig:scatter_redshift} shows the scatter plot between spectra and PAUS+COSMOS photo-$z$ (this work, top panel) and the \cosmos~photo-$z$ (bottom panel). Visually, both show a very tight concentration along the diagonal, with a small number of outliers seen in both catalogs. We note that a group of outliers confusing the $\OIII$ and $\Halpha$ lines at $\Delta_z\approx-0.24$ present in the \cosmos~catalogue get assigned the correct redshift in our new catalogue. Catastrophic outliers can occur for a number of different reasons, including failure of the model, missing or wrong SED templates, outliers in the photometry, and also outliers in the spectroscopic redshifts measurement. Both this work and the \cosmos~analysis share some of the above aspects, which explains why some outliers are present in both catalogues.

\subsection{Photometric redshift accuracy}

The previous section focused on the width of the $\Delta_z$ distribution to assess the average precision of the photo-$z$ point estimates in this catalogue. It is equally important to assess if the photo-$z$ estimates are also statistically unbiased, especially if they have to be used to calibrate the mean redshift of another sample very accurately.

\begin{figure}
	\includegraphics[width=\columnwidth]{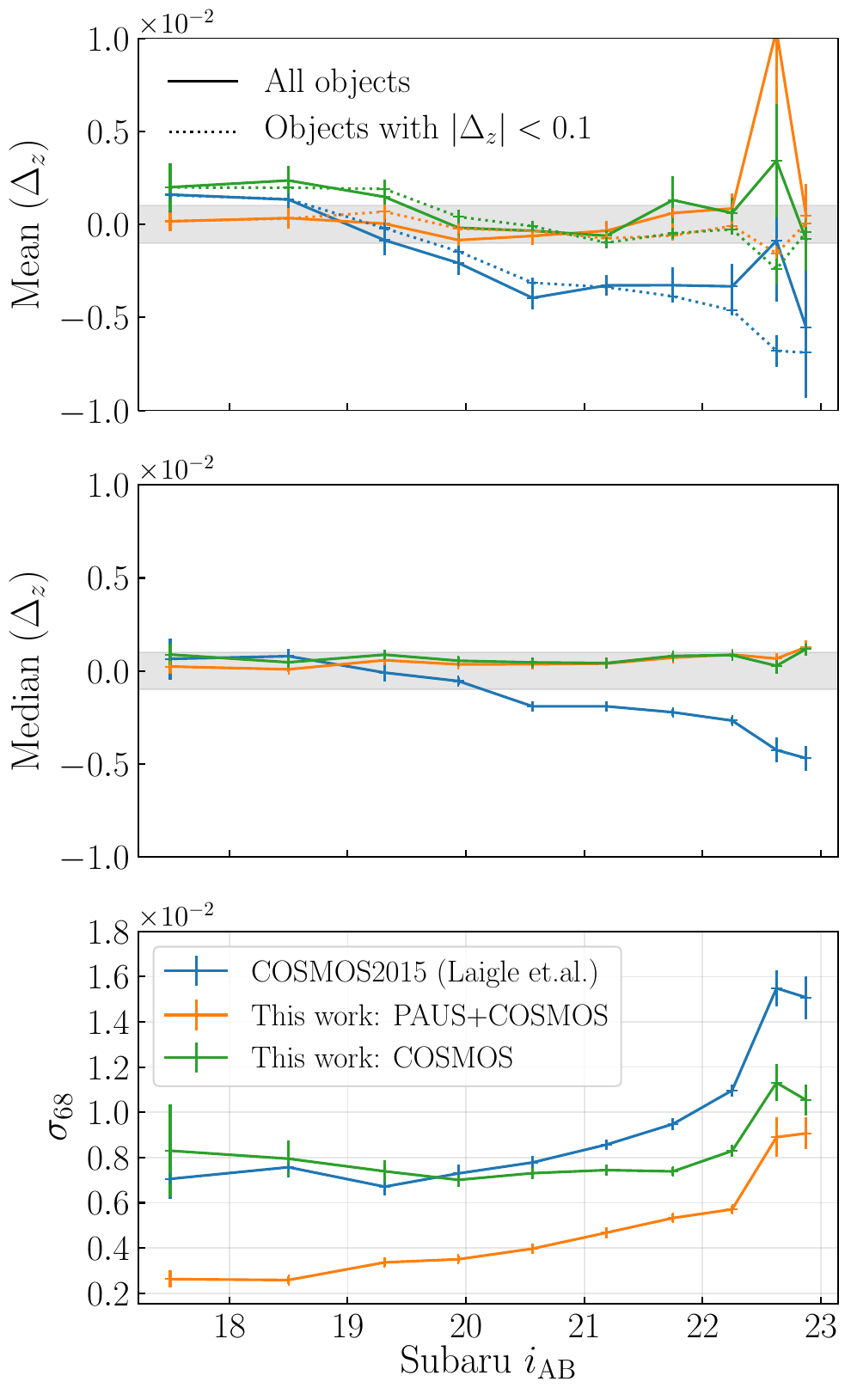}
    \caption{The mean (top panel), median (middle) and $\sigma_{68}$ (bottom) of the $\Delta_z$ distribution as a function of $\iab$ magnitude. The orange line shows the results from the fiducial calculation in this work, which includes photometry from PAUS and COSMOS surveys, while the green line is an additional calculation using only photometry from the COSMOS survey. For comparison, the blue lines show results using the photo-$z$ from \cosmos. We find the mean and median from this work to be statistically unbiased as a function of $\iab$ magnitude, as we find them consistent with $|\mathrm{mean}(\Delta_z)|\leq0.001$ and $|\mathrm{median}(\Delta_z)|\leq0.001$, which are shown as a shaded area in the two top panels. The bottom panel is equivalent to the top panel from Figure~\ref{fig:sigma68OR}, but adding the run which drops the PAUS photometry as a green line. The green lines across the panels show that most of the improvement in accuracy and a significant fraction of the improvement in precision at the faint end comes from the new methodology presented in this work, and not from the addition of the PAUS narrow band data. Most or all of the improvement in precision at $\iab \lesssim 21$ is achieved by including PAUS photometry. In the top panel, the dashed lines show the $\mathrm{mean}(\Delta_z)$ only for objects with $|\Delta_z|<0.1$, which shows the impact of the extreme outliers on this metric.}
    \label{fig:meanmedian_iband}
\end{figure}

\begin{figure}
	\includegraphics[width=\columnwidth]{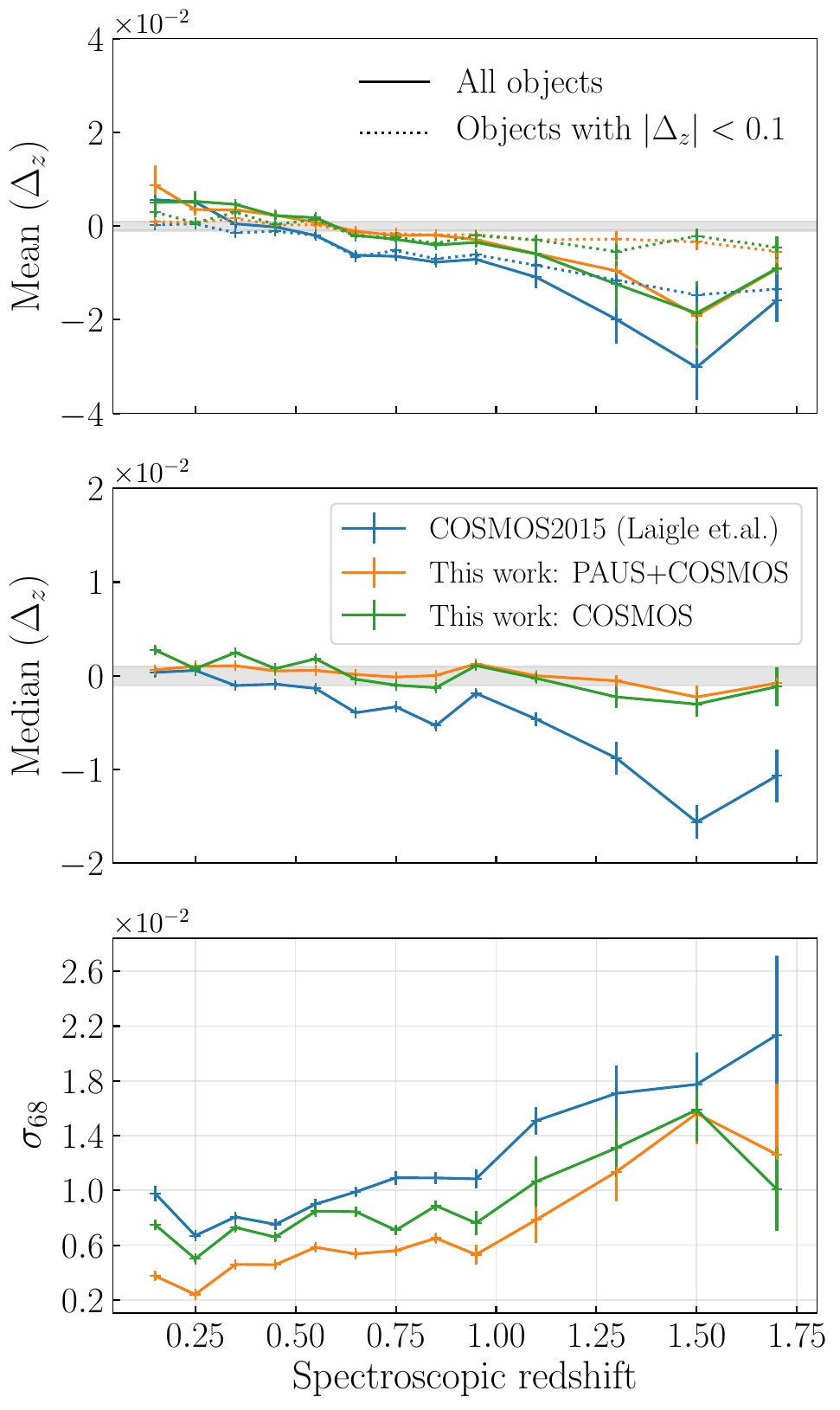}
    \caption{This figure is analogous to Figure~\ref{fig:meanmedian_iband}, showing the mean, median and $\sigma_{68}$ as a function of spectroscopic redshift. We find the median for the fiducial run in this work (orange line) to be consistent with $|\mathrm{median}(\Delta_z)|\leq0.001$ at all redshifts, showing that the center of the distribution is also unbiased as a function of redshift. The top panel, which shows the mean, shows significantly biased values, towards higher redshifts at low redshift, and towards lower redshift at higher redshifts. The dashed lines show the mean when removing extremely large outliers ($|\Delta_z|>0.1$), in which case the mean are pulled significantly closer to the 0.}
    \label{fig:meanmedian_zspec}
\end{figure}

\begin{figure}
	\includegraphics[width=\columnwidth]{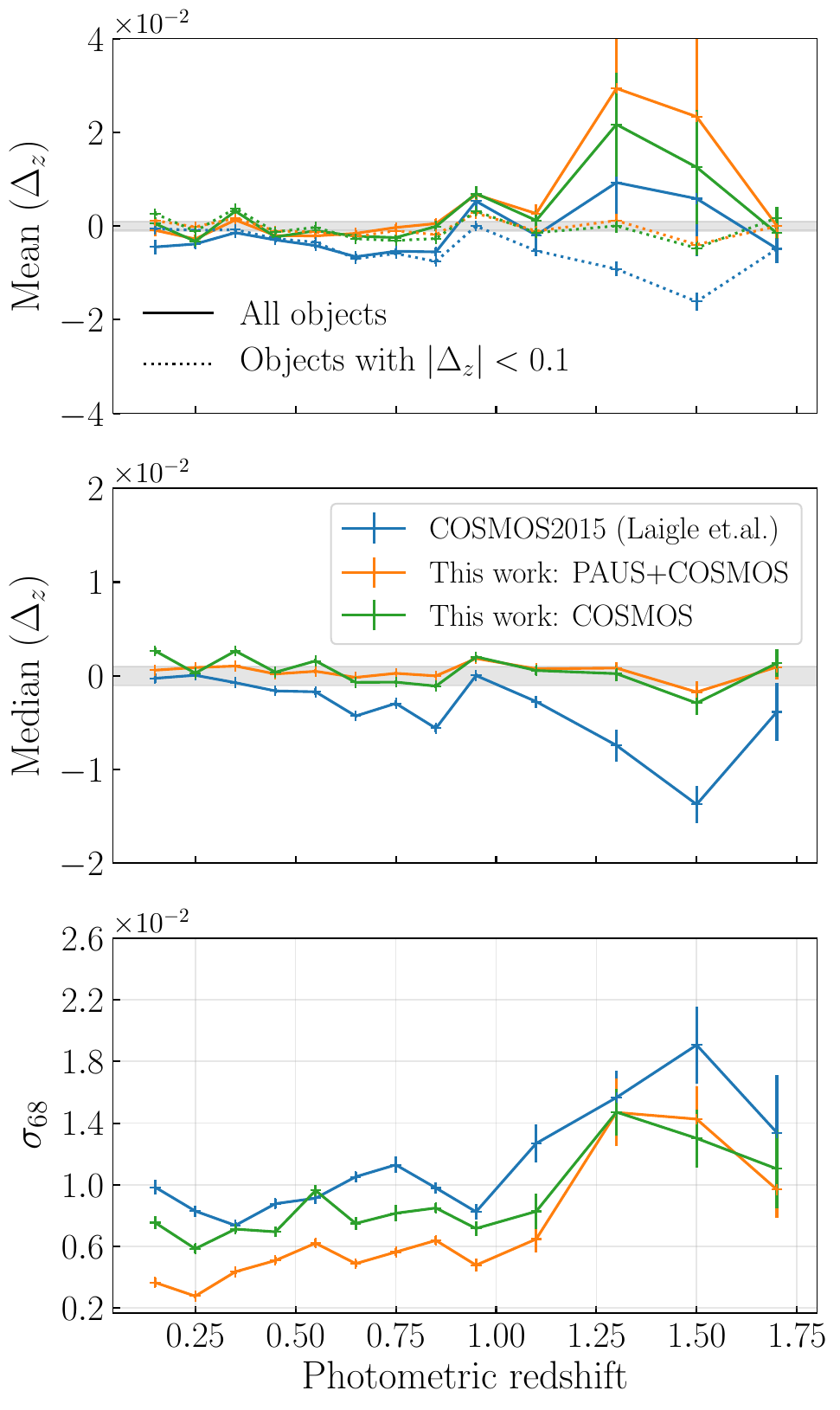}
    \caption{This figure is analogous to Figures~\ref{fig:meanmedian_iband}~and~\ref{fig:meanmedian_zspec}, showing the mean, median and $\sigma_{68}$ as a function of photometric redshift. We find the median for the fiducial run in this work (orange line) to be consistent with $|\mathrm{median}(\Delta_z)|\leq0.001$ at all redshifts, showing that the center of the distribution is also unbiased as a function of photo-$z$. The top panel, which shows the mean, shows more unbiased results than when binning as a function of spectroscopic redshift (Figure~\ref{fig:meanmedian_zspec}). The dashed lines show the mean when removing extremely large outliers ($|\Delta_z|>0.1$), in which case the mean becomes very consistent with $|\mathrm{mean}(\Delta_z)|\leq0.001$.}
    \label{fig:meanmedian_photoz}
\end{figure}

Figure~\ref{fig:meanmedian_iband} presents the mean, median and $\sigma_{68}$ of the $\Delta_z$ distribution as a function of Subaru $\iab$ magnitude. Figure~\ref{fig:meanmedian_zspec} presents the same variables but as a function of the spectroscopic redshift. While the mean is sensitive to outliers, the median should be more robust to them, which is why the bootstrap errorbars are smaller in the middle panels in both figures. We find the median of the $\Delta_z$ distribution to be consistent  within $|\mathrm{median}(\Delta_z)|\leq0.001$ (which is shown as a grey band) at all redshifts and magnitudes considered in this analysis (orange lines, middle panels), a threshold that we consider good enough to call these distributions statistically unbiased because it is much smaller than the central 68\% dispersion. In comparison, we find \cosmos~(blue lines) to be biased towards lower redshifts with respect to the spectroscopic sample, with a larger bias at higher redshift and fainter magnitudes. We find the mean from our work to also be unbiased at different magnitude bins (except for one bin), but we find a biased mean as a function of spectroscopic redshift. We find the catastrophic outlier fraction, objects with $|\Delta_z|>0.1$, to be mainly responsible for these biases in the mean, as shown by the dashed lines in the top panels of Figures~\ref{fig:meanmedian_iband} and \ref{fig:meanmedian_zspec}, which show the mean when removing these catastrophic outliers. Figure~\ref{fig:meanmedian_photoz} shows the mean, median and $\sigma_{68}$ as a function of photometric redshift. Similar to Figures~\ref{fig:meanmedian_iband} and \ref{fig:meanmedian_zspec} the median is unbiased, consistent with $|\mathrm{median}(\Delta_z)|\leq0.001$, as a function of photo-$z$. We find the mean to be more unbiased as a function of photo-$z$ than as a function spec-$z$, especially when removing the extreme outliers $|\Delta_z|>0.1$.

A systematic bias in the photo-$z$ estimates can lead to a bias in the cosmological inference of lensing surveys \citep[\textit{e.g.} see Figure 5 from][]{Salvato19}. In particular, \citet{KidsDEScomb} estimates the DES-Y1 source redshift distributions  to have a mean redshift about $0.01\sim0.05$ lower when estimated using \cosmos~instead of spectra. Here we also find biases towards lower redshift between \cosmos~ and spectra, although we find them to be lower than $\lesssim0.01$ in the magnitudes and redshift considered, and these get significantly reduced in our new catalogue. Understanding the biases found in \citet{KidsDEScomb} (importance of the faint end $\iab>23$, different spectroscopic samples, N(z) methodology) is beyond the scope of this work.

To understand which improvement comes from adding the narrow band photometry from PAUS and which comes from the new methodology, we have run the code excluding the PAUS photometry, using only the photometry coming from COSMOS. We show the mean, median and $\sigma_{68}$ for this study in the green lines in Figures~\ref{fig:meanmedian_iband}, \ref{fig:meanmedian_zspec} and \ref{fig:meanmedian_photoz}. We find comparable values for the mean and median between this run and the fiducial run which includes the PAUS photometry, indicating that the new methodology is responsible for obtaining the largely unbiased photo-$z$ estimates. Regarding the improvement in precision, the bottom panel of Figure~\ref{fig:meanmedian_iband} shows that part of the improvement in $\sigma_{68}$ at fainter magnitudes is explained by the new methodology, while most or all of the improvement in precision at $\iab \lesssim 21$ is achieved by including PAUS photometry.

\subsection{Performance of the complete sample}

\begin{figure}
	\includegraphics[width=\columnwidth]{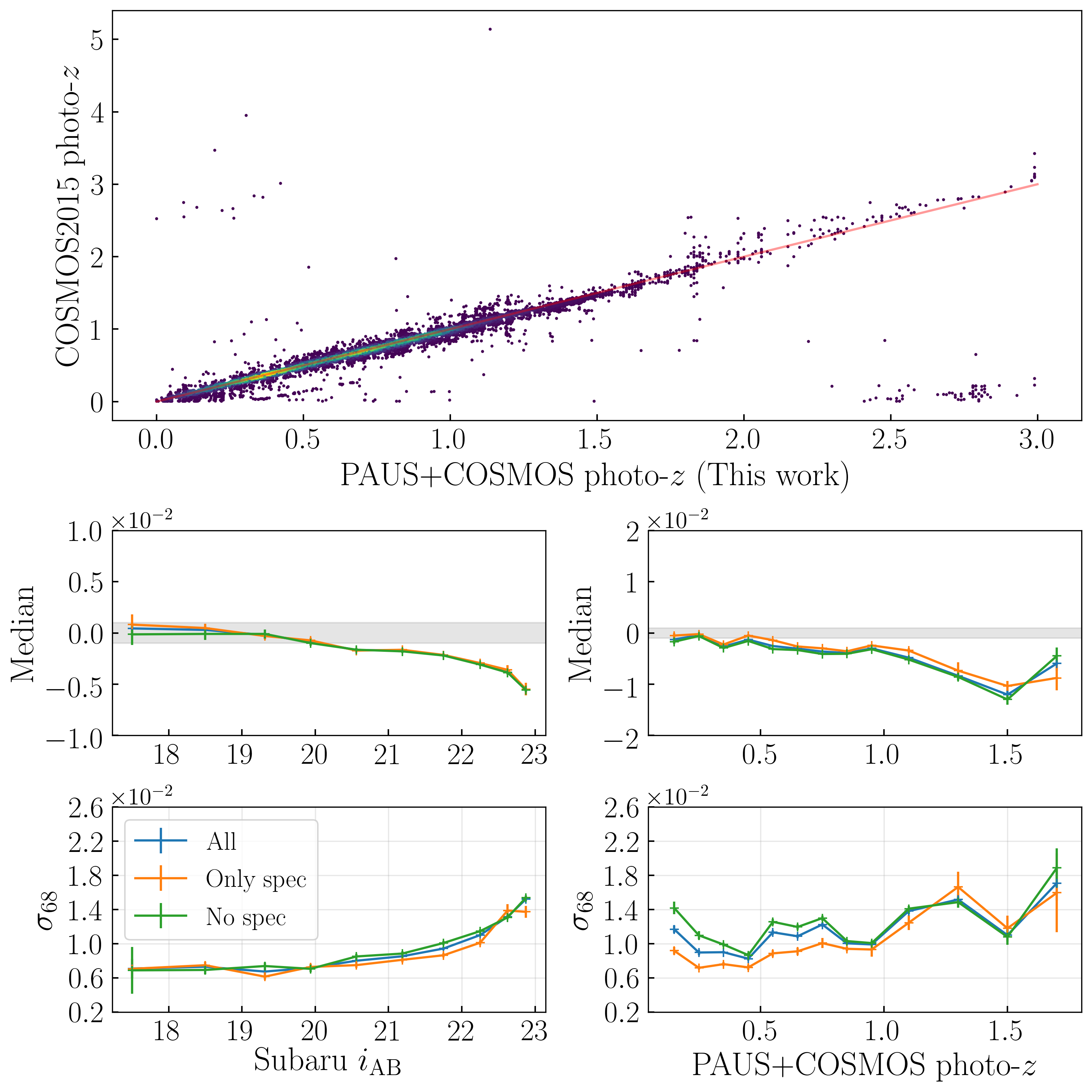}
    \caption{Performance of the complete sample. The top row shows the scatter plot between the photo-$z$ from PAUS+COSMOS and the photo-$z$ from \citep{COSMOS2015}. The middle and bottom row show the median and $\sigma_{68}$ of the quantity $\Delta_z \equiv (z_{\text{COSMOS2015}}-z_{\text{P+C}})/(1+z_{\text{P+C}})$, where $z_{\text{COSMOS2015}}$ and $z_{\text{P+C}}$ are the photo-$z$ from \cosmos~ and PAUS+COSMOS, as a function of the Subaru $i$-band (left panels) and the PAUS+COSMOS photo-$z$ (right panels). In the bottom panels, the blue lines show results for all objects in the catalog, the orange lines for the subset that has spectra, and the green lines for the subset that does not have spectra.}  
    \label{fig:fullsample}
\end{figure}

Figure~\ref{fig:fullsample} shows the performance of the full sample, including objects without spectroscopy. We do so via the quantity $\Delta_z \equiv (z_{\text{COSMOS2015}}-z_{\text{P+C}})/(1+z_{\text{P+C}})$, that compares the photo-$z$ from this work ($z_{\text{P+C}}$) and the photo-$z$ from \cosmos. We compare the value of the median and $\sigma_{68}$ of this quantity as a function of the Subaru $i$-band and the PAUS+COSMOS photo-$z$ for all objects and for two subsets: the one with spectroscopic redshift and the rest. We find the accuracy of the subset without spectra to be very similarly biased towards lower redshift, and to have a similar $\sigma_{68}$ as a function of magnitude ($\sim15\%$ higher), than the subset with spectra. The larger $\sigma_{68}$ as a function of photo-$z$ is likely explained by the different magnitude distribution, since the subset with spectra has a brighter magnitude distribution (see Figure~\ref{fig:spec_compl}).

In the top panel we observe again the outlier stripe at $\Delta_z\approx-0.24$, which shows a discrepancy between both catalogues at deciding between the $\OIII$ and $\Halpha$ lines. We also observe an outlier cloud with objects with low photo-$z$ according to \cosmos, that we find to be at high-$z$ ($z>2.3$) in our catalogue. Overlapping spectroscopic redshifts largely agree with our redshift assignment in the $\OIII$ v $\Halpha$ outlier stripe, but we do not find enough spectra to validate the existence of the low-z v high-z outlier cloud. However, note this outlier cloud follows the trend where the \cosmos~catalogue systematically assigns redshifts to lower possible values, and a similar cloud was also observed when using \textsc{LePhare} in simulated data \citep{Laigle2019}.

\section{Discussion and future work} \label{sec:discussion}

Our results from section~\ref{sec:results} demonstrate excellent photometric redshift average precision and accuracy across the redshifts and magnitudes considered in this work, which represents an important step towards a deep and complete redshift sample which is at the same time highly accurate and precise in redshift. There are several lines one can take to improve upon these results. One of them is to investigate the origin of the catastrophic outliers ($|\Delta_z|>0.1$). A rapid inspection by eye shows that a (small) fraction of these outliers could be explained by outlier photometry which has not been flagged during the data reduction process. Typically there are several exposures for the same filter and object, which are later \textit{coadded} either at the image level or later by averaging the flux measured in each exposure. Running the photo-$z$ algorithm directly on flux measurements from individual exposures could help reduce the impact of such photometric outliers. An alternative could be to develop a photometric outlier detection algorithm which finds and removes measurements which are likely to be spurious, but one would need to be careful to not remove real features (\textit{i.e.} emission lines). Finally, changing the Gaussian likelihood (Equation
~\ref{bayesevlikelihood}) to allow some measurements to be outliers could be an option, but it would require significant transformation of the algorithm.

The zero point re-calibration improves the photometric redshift estimation in template fitting techniques, as has been extensively pointed out in the literature (see section~\ref{sec:zeropoint_modeling}). The zero point factors from this work are shown in Figure~\ref{fig:offsets}, which are larger than the typical error of the measurements (statistical+systematic), and would degrade the photo-$z$ performance if the corrections were ignored. We have found that the iterative algorithm performs well using simulated data, finding that the correct answer is reached after convergence. However, we have found that if the colours are on average biased with respect to the colours in the sample, the iterative zero point re-calibration can introduce a very significant colour trend. In particular, in earlier stages of this work, when we set the emission line ratios to values from the literature instead of taking the median measured values from Figure~\ref{fig:beta_ratios}, the iterative algorithm would find a very large colour trend, lifting the flux from shorter wavelengths, which produced a statistically biased photo-$z$ estimation towards lower redshifts.

\begin{figure}
	\includegraphics[width=\columnwidth]{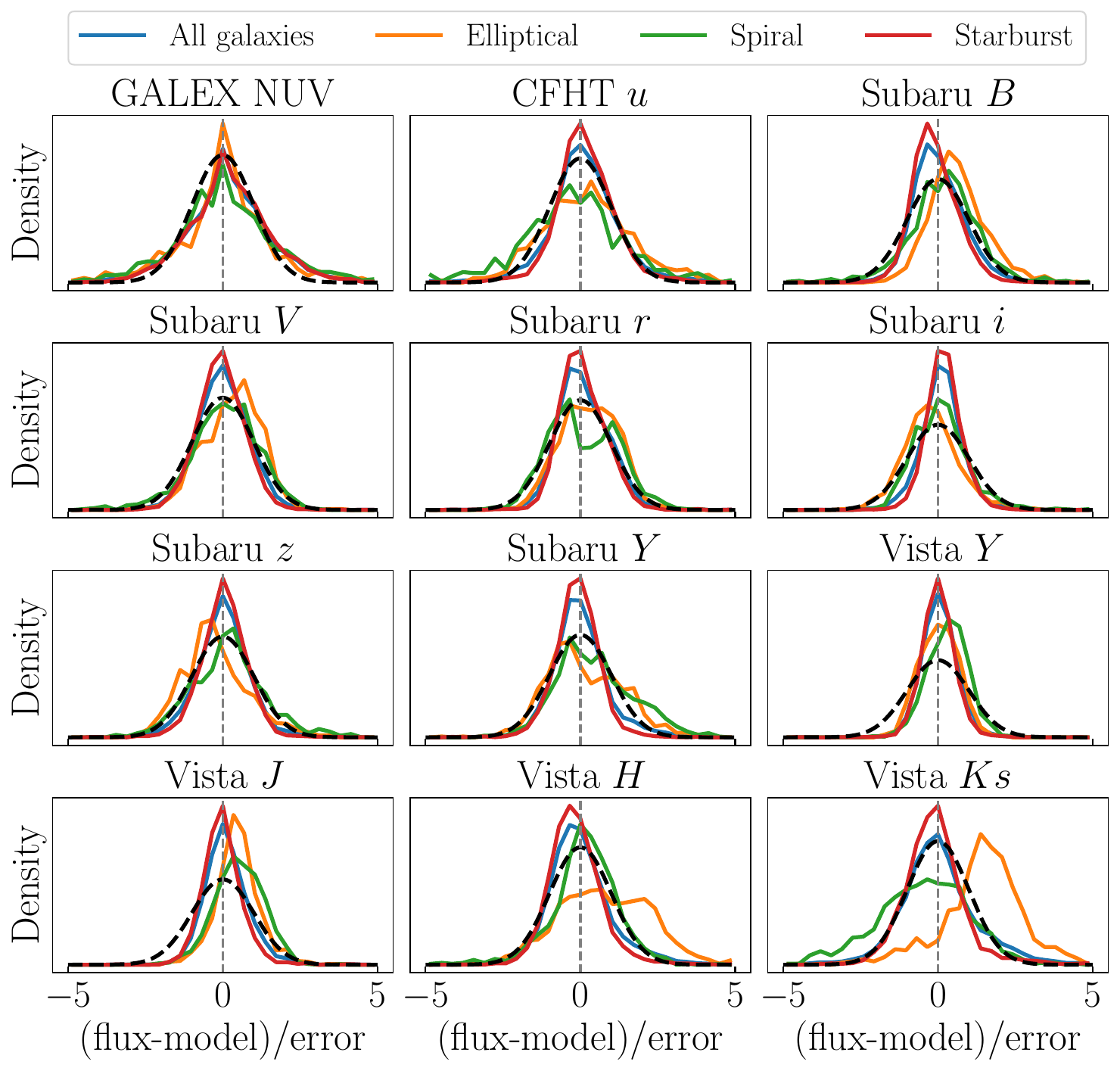}
    \caption{The density of residuals, defined as the difference between the measured flux and the most likely model, divided by the error (which includes both statistical and systematic error). Each panel shows one of the broad bands used in this work, and the residuals are shown for all galaxies and for three populations based on whether galaxies are classified to be elliptical, spiral or starburst according to which is the most likely model. The residuals show some significant offsets for elliptical galaxies and near infrared bands, which could indicate a problem with the models at larger wavelengths.}
    \label{fig:zp_pull}
\end{figure}

The zero point re-calibration essentially compares the measured fluxes to the most likely predicted fluxes from the model, for those objects which have a spectroscopic redshift. Figure~\ref{fig:zp_pull} shows the density of residual differences between the measured flux and the most likely model for different broad bands. The differences are weighted by the error, which includes both measured error and the systematic error per band discussed in section~\ref{sec:combined_catalog}. Furthermore, we show the density of residual differences for 3 subsets of the sample: whether the most likely model contains elliptical (orange lines), spiral (green) or starburst (red) templates. We find some interesting trends: while starburst galaxies seem to have unbiased residuals (although the errors appear to be overestimated), elliptical galaxies show a significant offset in the near infrared bands, where the models seem to lack flux. There are also some different trends for the spiral galaxies, which use the Prevot extinction law, as opposed to the Calzetti law in the starburst galaxies. 

The trends from Figure~\ref{fig:zp_pull} could indicate that there are missing templates; that the elliptical templates are wrong in the NIR; that there are many more starburst galaxies, which dominate the zero point recalibration and end up creating problems for elliptical galaxies; or that the Prevot law does not describe well on average the extinction of the spiral templates. We defer to future work the exploration of these possibilities, as well as studying combinations of elliptical templates with starburst templates (since the bulge and disk of a galaxy can have very different SEDs). An improvement of the models and templates would allow to decrease the systematic error per band and lead to a better exploitation of the statistical signal-to-noise.

\begin{figure}
	\includegraphics[width=\columnwidth]{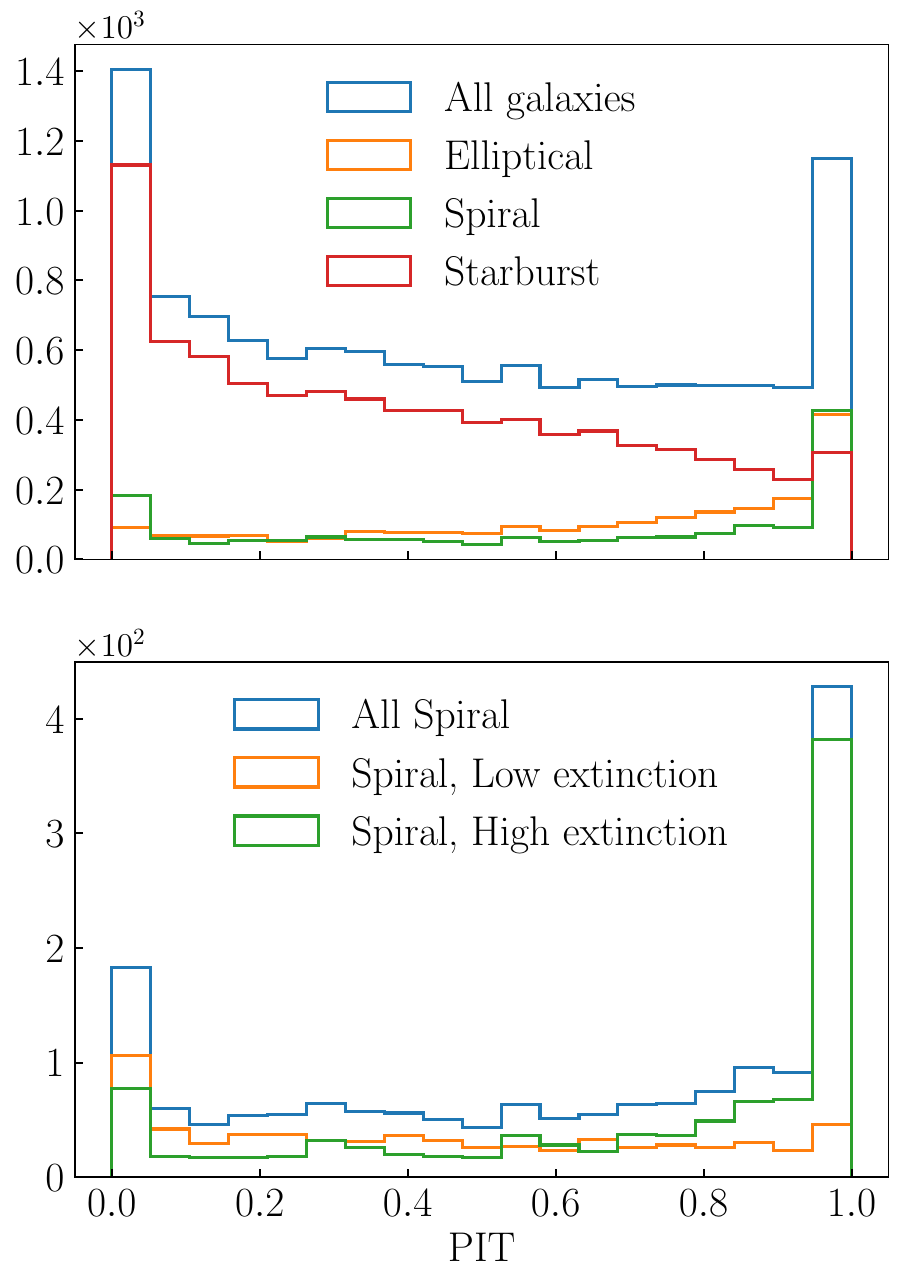}
    \caption{The distribution of the probability integral transform (PIT) values, defined as the cumulative distribution evaluated at the value of the spectroscopic redshift. The top panel shows the distribution for all objects with spectra, and for elliptical, spiral and starburst galaxies. The bottom panel show the PIT for spiral galaxies, and for spiral galaxies with a lower or higher internal dust extinction value.}
    \label{fig:PIT}
\end{figure}

Another interesting point of discussion is the redshift distribution of individual galaxies, which is required to be correct in many science applications (i.e. to correctly describe the uncertainties from the inference). In practice this is very rarely achieved because inaccuracies at describing the data will yield the wrong PDF. A common metric in the photo-$z$ literature for testing the quality of the PDF is the probability integral transform \citep[PIT, e.g.][]{Tanaka18,PAUSphotoz,deepz,Schmidt20}, which is defined as the cumulative distribution (CDF) evaluated at the spectroscopic redshift. If the PDFs were statistically correct an ensemble of PIT values should follow a uniform distribution. Figure~\ref{fig:PIT} shows the PIT values we obtain in this work. The top panel shows the PIT distribution for all galaxies (blue histogram), which shows an excess at both low and high PIT values. All catastrophic outliers ($|\Delta_z|>0.1$) live at the extremes of the PIT distribution, but also a few more objects with a reasonable photo-$z$ ($|\Delta_z|<0.01$) have PIT values closer than 0.01 to either PIT=0 or PIT=1 than they should. The top panel also shows the PIT distributions for elliptical, spiral and starburst galaxies, revealing that although the distribution for all galaxies looks relatively flat, there are some clear trends for different populations: red galaxies seem to have a PDF biased low in redshift, while blue galaxies seem to be biased high, even if the photo-$z$ (the mode of the pdf) is still very close to the spectroscopic redshift for all of them. Therefore, the PDFs can be deemed unreliable from a statistical point of view, which is a common feature in many photo-$z$ algorithms \citep[e.g. see Figure 2 from][]{Schmidt20}. However, it is important to point out that the redshift errors implied by the PIT distributions are smaller when the PDFs are narrow, as is the case for the PDFs we obtain. For example, we find $\mathrm{median}(\Delta_z)=0.0012$ for starburst galaxies and $\mathrm{median}(\Delta_z)=-0.0016$ for spiral and elliptical galaxies, which means that the errors in the PDFs seen in the PIT distribution do not translate into large redshift inaccuracies, as the PDFs are indeed narrow in redshift. 

The PDFs could be improved by extending the template baseline used in this work, using synthetic stellar population synthesis models to generate more adequate templates \citep[e.g.][]{conroy2009,fsps,Chaves2020}. We also find different PIT distribution for low or high extinction spiral galaxies (bottom panel), which could hint at a problem on how additional dust extinction is being modeled in spiral galaxies. One could also try to combine with AGN templates, or to add additional extinction to elliptical galaxies. A natural extension of the algorithm would be to allow more emission lines to be free, including adequate priors for different emission lines, instead of choosing some fixed values. The implicit redshift priors included in the priors for the continuum and emission line templates can also yield the wrong PIT distribution. We leave for a future study a detailed examination of these points.

Finally, PAUS has collected deeper data for a fraction of the COSMOS field, which will be a natural extension of this work towards improved redshift calibration samples for fainter objects.

\section{Summary and conclusions} \label{sec:conclusions}

We have presented a novel method to estimate photometric redshifts with multi-band photometric data. We have applied the method to data from the COSMOS field and we have assessed its performance comparing to available public spectroscopic redshifts. We have used a combination of 40 narrow band photometric filters from the PAUS survey and 26 broad, intermediate and narrow band filters from the COSMOS survey \citep[][COSMOS2015]{COSMOS2015} to estimate the most precise and accurate photometric redshifts available in the COSMOS field for objects with $\iab\leq23$. We have developed a new algorithm that models the galaxy SED using a linear combination of continuum and emission line templates and integrates over their possible different combinations using priors. The main primary results and
conclusions are:

\begin{enumerate} 
    \item We find a redshift precision of $\sigma_{68}(\Delta_z)\sim0.0057,~0.009$ at magnitude $\iab\sim22.5,~23$, respectively, which is over a factor $1.66\times$ tighter than previous results from \cosmos, with $\Delta_z\equiv(z_{\mathrm{photo}}-z_{\mathrm{spec}})/(1+z_{\mathrm{spec}})$. The precision gets much better at bright magnitudes, where we find $\sigma_{68}\sim0.0026$ at $\iab\sim18$ (see Figure~\ref{fig:sigma68OR}).
    \item We find the redshifts to be statistically unbiased, with the median of the $\Delta_z$ distribution consistent  within $|\mathrm{median}(\Delta_z)|\leq0.001$ at all redshifts and magnitudes considered in this analysis (see Figures~\ref{fig:deltaz_zcuts}, \ref{fig:meanmedian_iband}, \ref{fig:meanmedian_zspec},\ref{fig:meanmedian_photoz}).
    \item We measure different emission lines using the 66 photometric bands and a subsample with spectroscopic redshifts (see Figures~\ref{fig:beta_ratios}, \ref{fig:alpha_MUV_relation}). We use these measurements to build the emission line templates and calibrate a model to balance between the emission line and the continuum templates. This yields a galaxy model with more accurate colors, with produces a statistically unbiased redshift inference. 
    \item We make the redshift catalog publicly available through the \textsc{cosmohub} platform (see Appendix~\ref{app:cosmohub} for details of the catalog and how to download it).
    \end{enumerate}
    
The results from our work indicate that deeper PAUS data in the COSMOS field will significantly improve the performance and redshift precision of this photo-$z$ sample and potentially yield a redshift sample with $\sigma_{68}(\Delta_z)<0.01$ at $\iab>23$.

\section*{Acknowledgements}

AA would like to thank Mara Salvato for providing a compilation of public spectroscopic redshifts in the COSMOS field. AA would like to thank Clotilde Laigle and Olivier Ilbert for providing information on SED templates, filter transmission curves and filter zero points at early stages of this work. AA thanks Jonas Chaves-Montero, Andrew Hearin and Nesar Ramachandra for useful discussions on this work. Argonne National Laboratory's work was supported by the U.S. Department of Energy, Office of High Energy Physics. Argonne, a U.S. Department of Energy Office of Science Laboratory, is operated by UChicago Argonne LLC under contract no. DE-AC02-06CH11357. The PAU Survey is partially supported by MINECO under grants CSD2007-00060, AYA2015-71825, ESP2017-89838, PGC2018-094773, PGC2018-102021, SEV-2016-0588, SEV-2016-0597, MDM-2015-0509 and Juan de la Cierva fellowship and LACEGAL and EWC Marie Sklodowska-Curie grant No 734374 and no.776247 with ERDF funds from the EU Horizon 2020 Programme, some of which include ERDF funds from the European Union. IEEC and IFAE are partially funded by the CERCA and Beatriu de Pinos program of the Generalitat de Catalunya. Funding for PAUS has also been provided by Durham University (via the ERC StG DEGAS-259586), ETH Zurich, Leiden University (via ERC StG ADULT-279396 and Netherlands Organisation for Scientific Research (NWO) Vici grant 639.043.512), University College London and from the European Union's Horizon 2020 research and innovation programme under the grant agreement No 776247 EWC. The PAU data center is hosted by the Port d'Informaci\'o Cient\'ifica (PIC), maintained through a collaboration of CIEMAT and IFAE, with additional support from Universitat Aut\`onoma de Barcelona and ERDF. We acknowledge the PIC services department team for their support and fruitful discussions. This project has received funding from the European Union's Horizon 2020 research and innovation programme under grant agreement No 776247. H. Hildebrandt is supported by a Heisenberg grant of the Deutsche Forschungsgemeinschaft (Hi 1495/5-1) as well as an ERC Consolidator Grant (No. 770935). H. Hoekstra acknowledges support from Vici grant 639.043.512 from the Netherlands Organization for Scientific Research (NWO). GM acknowledges support from ST/P006744/1. CMB acknowledges support from ST/P000541/1 and ST/T000244/1. PN acknowledges support from ST/P000541/1 and ST/T000244/1. This project has received funding from the European Union's Horizon 2020 research and innovation programme under the Maria Sk\l{}odowska-Curie (grant agreement No 754510), the National Science Centre of Poland (grant UMO-2016/23/N/ST9/02963) and by the Spanish Ministry of Science and Innovation through Juan de la Cierva-formacion program ( reference FJC2018-038792-I).


\section*{Data availability}
\label{sec:data_availability}

The data underlying this article are available at
\begin{itemize}
\item Redshift catalog: redshift information for 40672 galaxies is publicly available at \url{https://cosmohub.pic.es/home} under \texttt{Catalogs} labelled as \texttt{PAUS+COSMOS photo-z catalog}.
\item PAUS images used in this work are available from the ING Observatory and also from the authors on request. The PAUS photometry used in this work is preliminary and is being used in other applications. We expect to make it available in upcoming publications.
\end{itemize}



\bibliographystyle{mnras}
\bibliography{biblio} 



\section*{Affiliations}
\noindent
{$^{1}$HEP Division, Argonne National Laboratory, Lemont, IL 60439
\\
$^{2}$Institute of Space Sciences (ICE, CSIC), Campus UAB, Carrer de Can Magrans, s/n, 08193 Barcelona, Spain \\
$^{3}$Institut d'Estudis Espacials de Catalunya (IEEC), E-08034 Barcelona, Spain \\
$^{4}$Institut de F\'{\i}sica d'Altes Energies (IFAE), The Barcelona Institute of Science and Technology, Campus UAB, 08193 Bellaterra (Barcelona), Spain  \\
$^{5}$Institute for Computational Cosmology (ICC), Department of Physics, Durham University, South Road, Durham DH1 3LE, UK\\
$^{6}$Institute for Data Science (IDAS), Durham University, South Road, Durham DH1 3LE, UK\\
$^{7}$CIEMAT, Centro de Investigaciones Energ\'eticas, Medioambientales y Tecnol\'ogicas, Avda. Complutenes 40, 28040 Madrid, Spain\\
$^{8}$Instituto de Fisica Teorica (IFT-UAM/CSIC), Universidad Autonoma de Madrid, 28049 Madrid, Spain\\
$^{9}$Ruhr-University Bochum, Astronomical Institute, German Centre for Cosmological Lensing, Universit\"{a}tsstr. 150, 44801 Bochum, Germany\\
$^{10}$Leiden Observatory, Leiden University, Niels Bohrweg 2, Leiden, The Netherlands\\
$^{11}$Department of Physics and Astronomy, University College London, Gower Street, London WC1E 6BT, UK\\
$^{12}$Centre for Extragalactic Astronomy (CEA), Department of Physics, Durham University, South Road, Durham DH1 3LE, UK\\
$^{13}$Instituci\'o Catalana de Recerca i Estudis Avan\c{c}ats (ICREA), 08010 Barcelona 
}


\appendix

\section{Redshift catalogue details} \label{app:cosmohub}

We make the catalogue publicly available through the \textsc{cosmohub} platform: \url{https://cosmohub.pic.es} \citep{cosmohub_carretero18,cosmohub_tallada20}. After registering, the catalog is available under \texttt{Catalogs} labelled as \texttt{PAUS+COSMOS photo-z catalog}. We provide one catalogue with a list of summary properties for each galaxy, which include the best photometric redshift and best model information and a few absolute magnitudes and line flux measurements derived at the best photo-$z$ and best model. Table~\ref{tab:cosmohub_table} shows a summary of the columns provided in the catalogue, and a description of each column. A second file containing the redshift distributions (Equation~\ref{pseudoposterior}) for all galaxies can be directly downloaded from the \texttt{Value Added Data} section. The redshift distribution is evaluated between redshift 0 and 3: in steps of 0.001 between redshift 0 and 1, in steps of 0.002 between redshift 1 and 1.5, and in steps of 0.01 between redshift 1.5 and 3.0. Note the different redshift step sizes, especially when integrating over the redshift distribution.

\begin{table*}
	\centering
	\caption{Description of the columns provided in the photo-$z$ catalogue}
	\label{tab:cosmohub_table}
	\begin{tabular}{cl} 
		\hline
		Column name & Description\\
		\hline
\texttt{ref\_id} & PAUdm reference ID\\
\texttt{I\_auto} & Auto $i$-band magnitude from \cite{COSMOSILBERT} cosmos photo-$z$ catalogue \\
\texttt{photoz} & Photo-$z$, defined as the mode of the p(z) of each object obtained from the templates\\
\texttt{ra} & Right ascention\\
\texttt{dec} & Declination\\
\texttt{nbands} & Number of bands used in the photo-$z$ code\\
\texttt{id\_laigle} & \cosmos~\citep{COSMOS2015} reference ID\\
\texttt{zspec\_mean} & Average spectroscopic redshift from several public redshift surveys (when available)\\
\texttt{zspec\_std} & Standard deviation spectroscopic redshift from several public redshift surveys (when available)\\
\texttt{best\_model} & Best model in the photo-$z$ code\\
\texttt{best\_extlaw} & Best extinction law in the photo-$z$ code (0: No extinction; 1: Prevot; 2: Calzetti; 3: Calzetti+Bump1; 4: Calzetti+Bump2)\\
\texttt{best\_continuum} & Best continuum template group ([0,16] see Table~\ref{tab:bayesev_models})\\
\texttt{best\_ebv} & Best extinction E(B-V) value (from 0 to 0.5, in steps of 0.05). See Equation~\ref{extinction}.\\
\texttt{best\_type} & Best galaxy type, based on \texttt{best\_continuum} (0: red ([0,12]), 1: green ([13,14]), 2: blue ([15,16]))\\
\texttt{MUV} & Absolute magnitude Galex NUV band. Corrected for internal galaxy extinction.\\
\texttt{MU} & Absolute magnitude CFHT $u$ band. Corrected for internal galaxy extinction.\\
\texttt{MR} & Absolute magnitude Subaru $r$ band. Corrected for internal galaxy extinction.\\
\texttt{MI} & Absolute magnitude Subaru $i$ band. Corrected for internal galaxy extinction.\\
\texttt{flux\_OII} & Flux for the OII line template in units of 1e-17 erg/s/cm2. Corrected for internal galaxy extinction.\\
\texttt{flux\_Hbeta} & Flux for the Hbeta line template in units of 1e-17 erg/s/cm2. Corrected for internal galaxy extinction.\\
\texttt{flux\_OIII} & Flux for the OIII line template in units of 1e-17 erg/s/cm2. Corrected for internal galaxy extinction.\\
\texttt{flux\_Halpha} & Flux for the Halpha line template in units of 1e-17 erg/s/cm2. Corrected for internal galaxy extinction.\\
\texttt{flux\_err\_OII} & Flux error for the OII line template in units of 1e-17 erg/s/cm2. Corrected for internal galaxy extinction.\\
\texttt{flux\_err\_Hbeta} & Flux error for the Hbeta line template in units of 1e-17 erg/s/cm2. Corrected for internal galaxy extinction.\\
\texttt{flux\_err\_OIII} & Flux error for the OIII line template in units of 1e-17 erg/s/cm2. Corrected for internal galaxy extinction.\\
\texttt{flux\_err\_Halpha} & Flux error for the Halpha line template in units of 1e-17 erg/s/cm2. Corrected for internal galaxy extinction.\\
\texttt{line\_coverage\_OII} & \makecell[l]{Maximum coverage of the OII flux line. Range is [0,1], with 0 being poorly covered and 1 great coverage. \\Measurements with poor coverage are unreliable and should be flagged.}\\
\texttt{line\_coverage\_Hbeta} & \makecell[l]{Maximum coverage of the Hbeta flux line. Range is [0,1], with 0 being poorly covered and 1 great coverage. \\Measurements with poor coverage are unreliable and should be flagged.}\\
\texttt{line\_coverage\_OIII} & \makecell[l]{Maximum coverage of the OIII flux line. Range is [0,1], with 0 being poorly covered and 1 great coverage. \\Measurements with poor coverage are unreliable and should be flagged.}\\
\texttt{line\_coverage\_Halpha} & \makecell[l]{Maximum coverage of the Halpha flux line. Range is [0,1], with 0 being poorly covered and 1 great coverage. \\Measurements with poor coverage are unreliable and should be flagged.}\\
\hline
	\end{tabular}
\end{table*}

\section{Combining heterogeneous photometry} \label{NB2BB}

Here we describe the method to produce a synthetic broad band flux from an overlapping set of narrow band filters. This synthetic flux can be used to estimate a factor that calibrates heterogeneous photometry (see Section~\ref{sec:combined_catalog}).

Let $W_{B}(\lambda)\equiv\lambda^{-1}T_{B}(\lambda)$ be a broad band filter that we want to express as a linear combination of 40 narrow band filters  $W_{N}(i,\lambda)\equiv\lambda^{-1}T_{N}(\lambda)$, where $T(\lambda)$ are the filter responses (in units of photon) and $i=1,\ldots,40$. One can find a solution by writing the coefficients $c(i)$ such that 
\begin{equation}\label{WBWN}
W_{B}(\lambda) = \sum_{i=1}^{i=40} c(i) ~ W_{N}(i,\lambda)
\end{equation}
where $\lambda$ is the wavelength which we will bin in integer values. We define
\begin{equation}
\begin{split}
\langle W_{B}W_{B}\rangle  \equiv& \int ~d\lambda ~W_{B}^2(\lambda) = \sum_{\lambda} ~W_{B}^2(\lambda) = 1;\\
\langle W_{N}W_{N} \rangle \equiv&  \int ~d\lambda ~W_{N}^2(\lambda) = \sum_{\lambda} ~W_{N}^2 (\lambda) = 1;
\end{split}
\end{equation}
which indicates the transmission curve's norm is normalized to unity. We can then define 40 elements of broad-narrow  projection vector:
\begin{equation}
BN(j)\equiv \langle W_{B} W_{N}(j)  \rangle =  \int d\lambda ~W_{B}(\lambda) W_{N}(j,\lambda)
\end{equation}
and a 40x40 narrow band overlap matrix
\begin{equation}
NN(i,j)\equiv \langle W_{N}(i) W_{N}(j)  \rangle =  \int d\lambda ~W_{N}(i,\lambda) W_{N}(j,\lambda) .
\end{equation}
If we multiply Eq.\ref{WBWN} by $W_N(j,\lambda)$  and then integrate we get
\begin{equation}
BN(j) = \sum_i ~c(i) ~NN(i,j)  .
\end{equation}
As $NN(i,j)$ is invertible, the unique solution is
\begin{equation}\label{cvalues}
\begin{split}
c(i) &= \sum_j NN^{-1}(i,j)  BN(j) \\
&= \sum_j \int d\lambda ~W_{B}(\lambda) W_{N}(j,\lambda) ~\left[\int d\lambda W_{N}(i,\lambda) W_{N}(j,\lambda) ~\right]^{-1} 
\end{split}
\end{equation}
We can use the $c(i)$ values from Eq. \ref{cvalues} to build a synthetic broad band flux from 40 narrow bands fluxes as
\begin{equation}
f^{BB}_{syn} \equiv \sum_{i=1}^{i=40} c(i) ~ f_i^{NB}
\end{equation}
Comparing the synthetic Subaru $r$-band with the measured broad band we estimate a factor that we can apply to bring two different photometric systems together.
 
\begin{figure}
\centering
\includegraphics[width=\columnwidth]{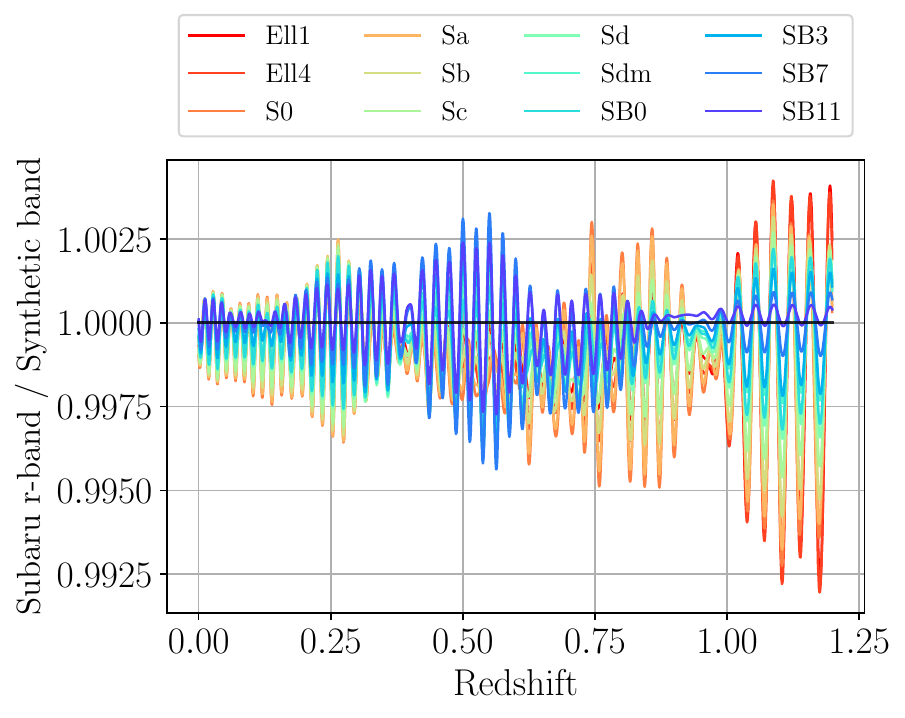}
\caption{Ratio between the Subaru r-band true flux and the synthetic flux obtained from the narrow band fluxes for different SED models (colour coded) at different redshifts (x-axis). The deviations from unity can be predicted and included in the SED modelling.}
\label{rsynthrationoem}
\end{figure}

Due to the particular shape and small overlap between different narrow band filters, the synthetic broad band flux will slightly differ from the true one. This effect can be predicted and included in the modelling of the galaxy SED. Fig. \ref{rsynthrationoem} shows the ratio between the synthetic broad band flux and the true broad band flux for different continuum SEDs at different redshifts, showing some oscillations around 1. Additionally, when narrow band measurements are missing, we extrapolate from the remaining narrow bands.

\section{Algorithms} \label{algorithm}

In this section we present details of the two photo-z algorithms that are used in this work.

\subsection{Minimization algorithm} \label{sec:algorithm_minimization}

To minimize the likelihood defined in Eq. \ref{chi2likelihood} we restrict to non-negative parameters to avoid unphysical solutions. We use \textsc{bcnz2} \citep{PAUSphotoz}, which have implemented a non-negative quadratic programming iterative algorithm based on \cite{Sha2007}, which defines
\begin{equation}
A_{xy} \equiv \sum_i \frac{t^{x}_{i} t^{y}_{i}}{\sigma^{2}(f_{i})}, \quad b_x \equiv \sum_i \frac{t^{x}_{i} f_i}{\sigma^{2}(f_{i})},
\end{equation}
for templates $x$ and $y$, where the sum is over the bands and ${f}_i$ and $\sigma(f_i)$ are the measured flux and flux error. The amplitudes $\alpha_x$ get updated iteratively using
\begin{equation}
m_x = \frac{b_x}{\sum_y A_{xy} \alpha_y}, \quad \alpha_x \rightarrow m_x \alpha_x
\end{equation}
In the implementation the minimum is estimated at the same time for a set of galaxies, for all the different redshift bins.

\subsection{Bayesian Evidence integral algorithm} \label{sec:algorithm_integral}

To compute the Bayesian evidence we need to integrate Eq. \ref{bayesev_modeling}. Let us redefine the following terms from Eq. \ref{bayesevlikelihood}, $f_{k}\equiv f_k / \sigma(f_k)$ and $M_{jk}\equiv t_{jk} / \sigma(f_k)$. Then we can write the exponent in Eq. \ref{bayesevlikelihood} as 
\begin{equation}
\begin{split}
\sum_k^{d} \left(f_k - \sum_j^{n_i} \alpha_j\,M_{jk}\right)^2 &= \sum_k^{d} f_k^2 - 2 \sum_k^{d} \sum_j^{n_i}  f_k \, \alpha_j\,M_{jk} +\\
& + \sum_k^{d} \sum_j^{n_i} \sum_\ell^{n_i} \alpha_j \alpha_\ell \, M_{jk} M_{\ell k}\\
&= \pmb{f}\cdot \pmb{f} -2 \pmb{B}\cdot \pmb{\alpha} + \vec{\alpha}^\intercal  \hat{A}\,  \pmb{\alpha}
\end{split}
\end{equation} 
where in the second step we have defined the matrix $A_{j\ell} \equiv \sum_k^{d} M_{jk} M_{\ell k}$ and vector $B_j \equiv \sum_k^{d} f_k M_{jk}$. Then, Eq. \ref{bayesevlikelihood} becomes
\begin{equation}\label{likelihood2}
P(\pmb{f}|\pmb{\alpha}, z,M) = \frac{\exp(-\frac{1}{2}\pmb{f}\cdot \pmb{f})}{\sqrt{(2\pi)^{d}}\prod_k \sigma(f_k)} \exp\left[-\frac{1}{2}\pmb{\alpha}^\intercal  \hat{A}\,  \pmb{\alpha}  + \pmb{B}\cdot \pmb{\alpha}  \right]
\end{equation}
and then Eq. \ref{bayesev_modeling} becomes
\begin{equation} \label{integral2}
\begin{split}
P(\pmb{f}|z,M)  \propto& \int_0^{\Delta_1} \ldots \int_0^{\Delta_{n_i}} \, \exp\left[-\frac{1}{2}\pmb{\alpha}^\intercal  \hat{A}\,  \pmb{\alpha}  + \pmb{B}\cdot \pmb{\alpha}  \right]\\
&\times P(\pmb{\alpha} | z,M)\, d\pmb{\alpha}
\end{split}
\end{equation}
where we have dropped the constants that do not depend on the model. Eq. \ref{integral2} means integrating a prior function times a miscentered  multivariate normal distribution, which we can re-center with the following change of variable
\begin{equation} \label{integral3}
\begin{split}
P(\pmb{f}|z,M)  \propto& \int_0^{\pmb{\Delta}} \exp\left[-\frac{1}{2}\pmb{\alpha}^\intercal  \hat{A}\,  \pmb{\alpha}  + \pmb{B}\cdot \pmb{\alpha}  \right] P(\pmb{\alpha} | z,M)d\pmb{\alpha} \\
=& \exp(\frac{1}{2}\pmb{\mu}\cdot\pmb{B}) \int_0^{\pmb{\Delta}}  \exp\left[-\frac{1}{2}(\pmb{\alpha}-\pmb{\mu})^\intercal  \hat{A}\,  (\pmb{\alpha}-\pmb{\mu}) \right] \\
&\times P(\pmb{\alpha} | z,M) d\pmb{\alpha} \\
=& \exp(\frac{1}{2}\pmb{\mu}\cdot\pmb{B}) \int_{-\pmb{\mu}}^{\pmb{\Delta}-\pmb{\mu}} \exp\left[-\frac{1}{2}\pmb{\theta}^\intercal  \hat{A}\,  \pmb{\theta} \right] \\
&\times P(\pmb{\theta} | z,M)d\pmb{\theta} \\
\end{split}
\end{equation} 
where $\pmb{\theta} = (\pmb{\alpha}-\pmb{\mu})$ and $\pmb{\mu} = \hat{A}^{-1}\pmb{B}$ corresponds to the peak vector of the distribution (unconstrained maximum likelihood point). We implement the algorithm from (\citealt{Genz}), which consists in three transformations that make the numerical integration of Eq. \ref{integral3} more efficient. Following \cite{Genz}, we define the integral of a Gaussian as function $F$ of its integration limits $\pmb{a}$ and $\pmb{b}$,
\begin{equation} \label{fab}
F(\pmb{a},\, \pmb{b}) \equiv \frac{1}{\sqrt{|\hat{A}^{-1}|(2\pi)^n}} \int_{a_1}^{b_1} \ldots \int_{a_n}^{b_n} e^{-\frac{1}{2}\pmb{\theta}^\intercal  \hat{A}\,  \pmb{\theta} } p(\pmb{\theta}) d \pmb{\theta}.
\end{equation}
Then our integral becomes
\begin{equation} \label{final_integral}
P(\pmb{f}|z,M) \propto \exp(\frac{1}{2}\pmb{\mu}\cdot\pmb{B})\sqrt{|\hat{A}^{-1}|(2\pi)^{n_i}} \, F(\{-\pmb{\mu}\},\, \{\pmb{\Delta}-\pmb{\mu}\})
\end{equation} 
The transformations and code for estimating $F(\pmb{a},\, \pmb{b})$ in Eq.~\ref{fab} are described in the following section Appendix~\ref{sec:genz_algorithm}. The total volume under the multivariate Gaussian (Eq.~\ref{total_volume_mvn}) can be analytically derived as
\begin{equation} \label{total_volume_mvn}
\exp\left(\frac{1}{2}\pmb{\mu}\cdot\pmb{B}\right)\sqrt{|\hat{A}^{-1}|(2\pi)^{n_i}} = \int_{-\infty}^{\infty} \exp\left[-\frac{1}{2}\pmb{\alpha}^\intercal  \hat{A}\,  \pmb{\alpha}  + \pmb{B}\cdot \pmb{\alpha}  \right] d^{n_i} \alpha
\end{equation} 
which is a upper bound on the value of Eq.~\ref{bayesev_modeling}. The code takes advantage of this upper bound and avoids computing the more expensive integral when it has a value smaller than $500\times$ the current largest probability value. This makes the code runtime scale better for a larger redshift grid, or more models $M$.

\subsection{Gaussian integral} \label{sec:genz_algorithm}

In this subsection we reproduce the relevant details of the algorithm from \cite{Genz} used for the integration of Eq. \ref{fab} using 3 transformations. The first transformation will use the Cholesky decomposition of $\hat{C}\hat{C}^\intercal = \hat{A}^{-1}$, $\vec{\theta} = \hat{C}\vec{y}$. In this case $\vec{\theta}^\intercal  \hat{A}\,  \vec{\theta} = \vec{y}^\intercal  \vec{y}$ and $d \vec{\theta} = |\hat{A}^{-1}|^{\frac{1}{2}} d\vec{y}$. Note also how for this decomposition the new integration limits will be:
\begin{equation}
\begin{split}
\vec{a} \leq \vec{\theta} &= \hat{C}\vec{y} \leq  \vec{b}  \\
a_i \leq \theta_i &= \sum_j^n C_{ij} y_j \leq  b_i \\
\tilde{a}_i \equiv (a_i - \sum_{j\neq i} C_{ij} y_j)/C_{ii} \leq\,  &y_i \leq  (b_i -\sum_{j\neq i} C_{ij} y_j)/C_{ii} \equiv \tilde{b}_i  \\
\end{split}
\end{equation} 
Note that if $C$ is the lower triangular, $C_{ij} = 0$ for $i<j$. Hence,
\begin{equation}
\begin{split}
F(\vec{a},\, \vec{b}) =& \frac{1}{\sqrt{(2\pi)^n}} \int_{\tilde{a}_1}^{\tilde{b}_1} e^{-\frac{y^2_1}{2} }  \int_{\tilde{a}_2(y_1)}^{\tilde{b}_2(y_1)} e^{-\frac{y^2_2}{2} }  \ldots \\
& \times\ldots \int_{\tilde{a}_n(y_1,\ldots,y_{n-1})}^{\tilde{b}_n(y_1,\ldots,y_{n-1})} e^{-\frac{y^2_{n}}{2} } d \vec{y}.    
\end{split}
\end{equation} 
Now, the second transformation, $y_i = \Phi^{-1}(z_i)$, where
\begin{equation}
\begin{split}
\Phi(y) &= \frac{1}{\sqrt{2\pi}} \int_{-\infty}^y e^{-\frac{1}{2}\theta^2 } d \theta = \frac{1}{2}\left(\erf\left(\frac{y}{\sqrt{2}}\right)+1\right) \\
\Phi^{-1}(z_i) &= \sqrt{2}\, \erf^{-1} (2z_i -1)
\end{split}
\end{equation} 
then $dz_i = \frac{1}{\sqrt{2\pi}} e^{-\frac{1}{2}y_i^2 } dy_i$, so
\begin{equation}
\begin{split}
 F(\vec{a},\, \vec{b}) =& \int_{\Phi(\tilde{a}_1)}^{\Phi(\tilde{b}_1)}   \int_{\Phi(\tilde{a}_2(\Phi^{-1}(z_1)))}^{\Phi(\tilde{b}_2(\Phi^{-1}(z_1)))}   \ldots \\
 &\times\ldots \int_{\Phi(\tilde{a}_n(\Phi^{-1}(z_1),\ldots,\Phi^{-1}(z_{n-1})))}^{\Phi(\tilde{b}_n(\Phi^{-1}(z_1),\ldots,\Phi^{-1}(z_{n-1})))}  d \vec{z}.   
\end{split}
\end{equation} 
Finally, the third transformation, $z_i = d_i + w_i(e_i-d_i)$,
\begin{equation}
F(\vec{a},\, \vec{b}) = (e_1-d_1)\int_0^1 (e_2-d_2) \ldots \int_0^1 (e_n-d_n)  \int_0^1 d\vec{w}
\end{equation} 
where
\begin{equation}
\begin{split}
d_i &= \Phi\left( \left[a_i - \sum_{j\neq i} C_{ij} \Phi^{-1}(d_j + w_j(e_j-d_j))\right]/C_{ii} \right)\\
e_i &= \Phi\left( \left[b_i - \sum_{j\neq i} C_{ij} \Phi^{-1}(d_j + w_j(e_j-d_j))\right]/C_{ii} \right)
\end{split}
\end{equation} 
Note how for some parameters where either $a_i = -\infty$ or $b_i = \infty$ then $d_i=0$ and $e_i=1$, respectfully. Now we describe the algorithm to find the integral \citep[copied from][]{Genz}.
\begin{enumerate}
\item Input $\vec{a}$, $\vec{b}$, $\hat{A}$, $N_{max}$.
\item Compute Cholesky decomposition $\hat{C}$ for $A^{-1}$.
\item Initialize $\mathrm{Intsum}=0$, $N=0$, $d_1 = \Phi(a_1/C_{1,1})$, $e_1 = \Phi(b_1/C_{1,1})$, $f_1 = e_1 - d_1$.
\item Repeat until $N=N_{max}$
\begin{enumerate}
\item Generate uniform random $w_1, w_2, \ldots, w_{m-1}\in[0,1]$.
\item For $i=2,3,\ldots, m$ set $y_{i-1} = \Phi^{-1}(d_{i-1}+w_{i-1}(e_{i-1}-d_{i-1}))$, $d_i = \Phi((a_i-\sum_{j=1}^{i-1}y_j C_{ij})/C_{ii})$, $e_i = \Phi((b_i-\sum_{j=1}^{i-1}y_j C_{ij})/C_{ii})$, $f_i = (e_i-d_i)f_{i-1}$.
\item Set $N=N+1$, $\delta = (f_m-\mathrm{Intsum})/N$, $\mathrm{Intsum}=\mathrm{Intsum}+\delta$.
\end{enumerate}
\item Output = $\mathrm{Intsum}$
\end{enumerate}
When there is a function $p(\pmb{\theta})$ multiplying the Gaussian inside the integral which depends on the parameters $\pmb{\theta}$ one needs to generate $w_m$ in step $(iv)(a)$, then $y_m$ in step $(iv)(b)$ and an additional substep to explicitly compute $\pmb{\theta}=C\pmb{y}$. Finally in step $(iv)(c)$ replace $f_m\rightarrow f_m\,p(\pmb{\theta})$.

\section{Systematic offsets} \label{app:sysoffs}

\begin{figure}
	\includegraphics[width=\columnwidth]{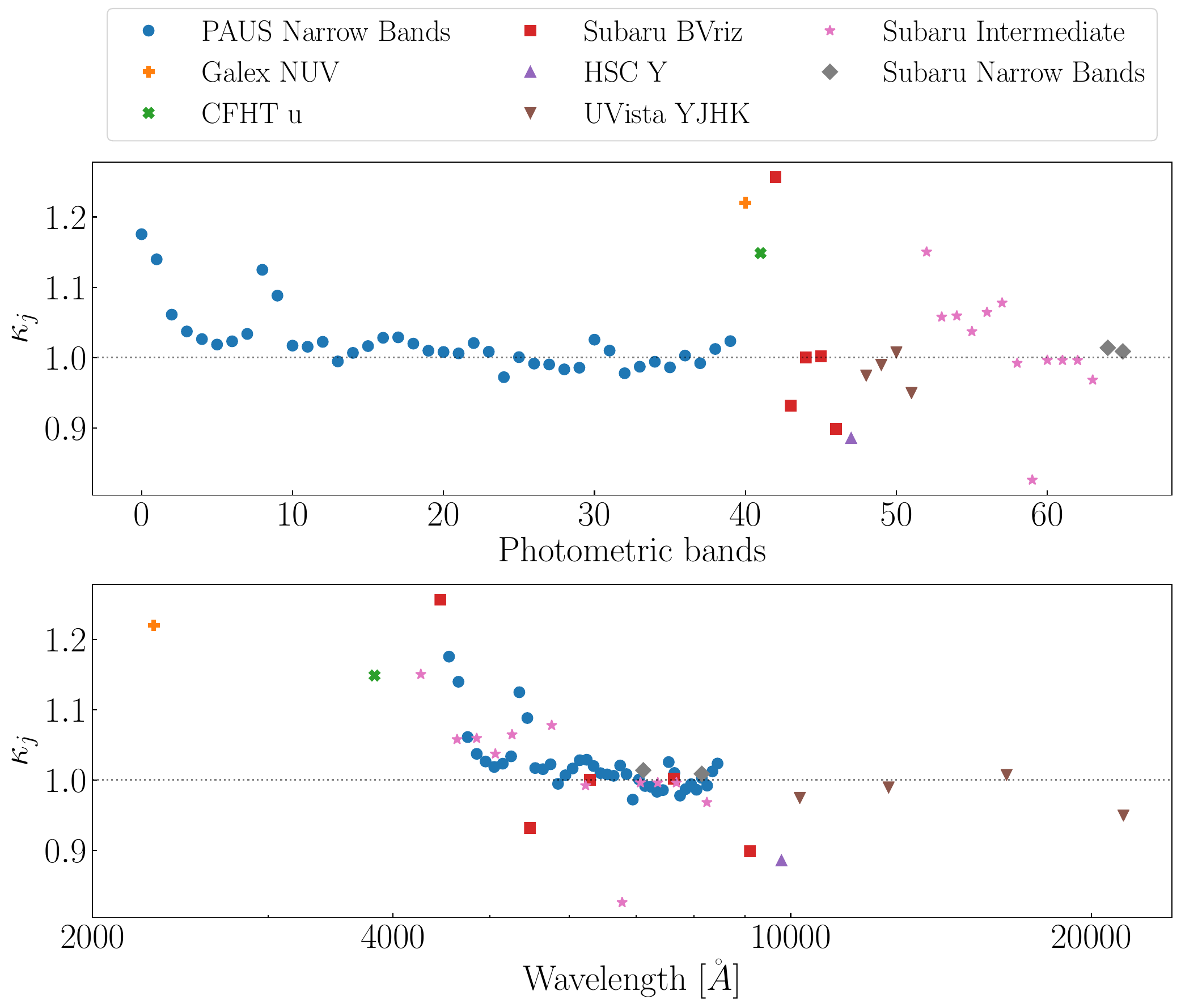}
    \caption{Systematic flux factors (values from Table~\ref{tab:offsets}) color coded in blocks of bands. The bottom panel shows the zero point of each band at the mean wavelenght of the band.}
    \label{fig:offsets}
\end{figure}

Here we show the systematic offsets described in section~\ref{sec:zeropoint_modeling}. Table~\ref{tab:offsets} and Figure~\ref{fig:offsets} show the value of the offsets we obtain for each of the bands we use in this work.

\begin{table}
\begin{tabular}{ c c c  }
\hline NB455 $\rightarrow$ 1.175 & NB465 $\rightarrow$ 1.140 & NB475 $\rightarrow$ 1.061 \\
 NB485 $\rightarrow$ 1.037 & NB495 $\rightarrow$ 1.026 & NB505 $\rightarrow$ 1.018 \\
  NB515 $\rightarrow$ 1.023 & NB525 $\rightarrow$ 1.034 & NB535 $\rightarrow$ 1.125 \\
 NB545 $\rightarrow$ 1.088 & NB555 $\rightarrow$ 1.017 & NB565 $\rightarrow$ 1.015 \\
 NB575 $\rightarrow$ 1.022 & NB585 $\rightarrow$ 0.994 & NB595 $\rightarrow$ 1.007 \\
 NB605 $\rightarrow$ 1.016 & NB615 $\rightarrow$ 1.028 &NB625 $\rightarrow$ 1.029 \\
 NB635 $\rightarrow$ 1.020 & NB645 $\rightarrow$ 1.010 & NB655 $\rightarrow$ 1.008 \\
  NB665 $\rightarrow$ 1.006 &NB675 $\rightarrow$ 1.021 & NB685 $\rightarrow$ 1.008 \\
 NB695 $\rightarrow$ 0.972 & NB705 $\rightarrow$ 1.001 & NB715 $\rightarrow$ 0.991 \\
 NB725 $\rightarrow$ 0.990 & NB735 $\rightarrow$ 0.983 & NB745 $\rightarrow$ 0.985 \\
 NB755 $\rightarrow$ 1.025 & NB765 $\rightarrow$ 1.010 & NB775 $\rightarrow$ 0.978 \\
 NB785 $\rightarrow$ 0.987 & NB795 $\rightarrow$ 0.994 & NB805 $\rightarrow$ 0.986 \\
 NB815 $\rightarrow$ 1.003 & NB825 $\rightarrow$ 0.992 & NB835 $\rightarrow$ 1.012 \\
 NB845 $\rightarrow$ 1.023 & Galex NUV $\rightarrow$ 1.22 & CFHT u $\rightarrow$ 1.148 \\
 Subaru B $\rightarrow$ 1.256 & Subaru V $\rightarrow$ 0.932 & Subaru r  $\rightarrow$ 1.000 \\
 Subaru i $\rightarrow$ 1.001 & Subaru z $\rightarrow$ 0.898 & HSC y $\rightarrow$ 0.886 \\
 UVista Y $\rightarrow$ 0.974 & UVista J $\rightarrow$ 0.989 & UVista H $\rightarrow$ 1.007 \\
 UVista K $\rightarrow$ 0.950 & Sub IA427 $\rightarrow$ 1.150 & Sub IA464 $\rightarrow$ 1.058 \\
 Sub IA484 $\rightarrow$ 1.059 & Sub IA505 $\rightarrow$ 1.037 & Sub IA527 $\rightarrow$ 1.064 \\
 Sub IA574 $\rightarrow$ 1.078 & Sub IA624 $\rightarrow$ 0.992 & Sub IA679 $\rightarrow$ 0.826 \\
 Sub IA709 $\rightarrow$ 0.996 & Sub IA738 $\rightarrow$ 0.996 & Sub IA767 $\rightarrow$ 0.996 \\
 Sub IA827 $\rightarrow$ 0.968 & Sub NB711 $\rightarrow$ 1.014 & Sub NB711 $\rightarrow$ 1.009 \\ \hline
 \end{tabular}
\caption{Systematic flux factors, $\kappa_j$ in Equation~\ref{zeropopint_kappa}, (the flux version of systematic magnitude offsets) measured in this work to make the colors predicted by the models and the ones measured in data more similar. A subsample of objects with spectroscopic redshift available is used (see section~\ref{sec:zeropoint_modeling} for more details).}
\label{tab:offsets}
\end{table}

\section{Model prior} \label{app:popprior}

This section describes how we compute the population prior on the models $p(M)$ described in section~\ref{sec:population_prior} and Equation~\ref{pseudoposterior}. We will use 10 broad bands: the CFHT $u$ band, the Subaru B, V, $r$, $i$, $z$ bands and the UltraVista Y,J,H,K bands. These 10 bands constitute a 9 dimensional color space $\pmb{C}$. We write
\begin{equation} \label{prior_pop_colors}
    p(M) = \int  p(M,z,\pmb{C}) \,dz\,d\pmb{C} = \int  p(M,z|\pmb{C})p(\pmb{C}) \,dz\,d\pmb{C}
\end{equation}
where $p(\pmb{C})$ is the distribution of colors, which we estimate from the observed color distribution in the data. Using Bayes theorem
\begin{equation}
    p(M,z|\pmb{C}) \propto p(\pmb{C}|M,z)p(M,z)
\end{equation}
Note $p(M,z)$ is unknown, so we assume here it is a uniform distribution, and we update this with the color space information $p(\pmb{C})$ that we observe. We obtain $p(\pmb{C}|M,z)$ by integrating over the colors each model can produce at each redshift
\begin{equation}
\begin{split}
    p(\pmb{C}|M,z) &= \int p(\pmb{C},\pmb{\alpha}|M,z) \, d\pmb{\alpha} \\
    &=\int p(\pmb{C}|\pmb{\alpha},M,z)p(\pmb{\alpha}|M,z) \, d\pmb{\alpha}
\end{split}
\end{equation}
where the color $p(\pmb{C}|\pmb{\alpha},M,z)$ is completely specified for each $(\pmb{\alpha},M,z)$, and $p(\pmb{\alpha}|M,z)$ is the prior from Equation~\ref{prior_templates}. 

In practice, we discretize the color space $\pmb{C}$ using a self organizing map (or SOM), which is an unsupervised machine learning technique which projects data from a high dimensional space into a lower dimensional grid. We refer to each unit of the grid as an SOM cell. The process preserves the topology of the higher dimensional data, which means that objects that were close in the original space will also be closer in the lower dimensional grid. We will use the same algorithm described in \cite{Masters2015}, which we implement to produce a two dimensional SOM. 

We estimate $p(M,z|\pmb{C})$ by randomly drawing values of $(\pmb{\alpha},M,z)$ from $p(\pmb{\alpha},M,z)$ and assigning them to an SOM cell. The distribution $p(\pmb{C})$ is estimated from the number counts of observed colors in each SOM cell. Finally we sum over cells and redshift using Equation~\ref{prior_pop_colors}. One can further refine this prior by allowing it to also depend on the observed Subaru $i$ band magnitude $m_i$ of each galaxy as
\begin{equation} \label{prior_pop_colors_mag}
\begin{split}
    p(M|m_i) &= \int  p(M,z,\pmb{C}|m_i) \,dz\,d\pmb{C} \\
    &= \int  p(M,z|\pmb{C},m_i)p(\pmb{C}|m_i) \,dz\,d\pmb{C} \\
    &\approx \int  p(M,z|\pmb{C})p(\pmb{C}|m_i) \,dz\,d\pmb{C}    
\end{split}
\end{equation}
In practice, the only change between Equation~\ref{prior_pop_colors} and Equation~\ref{prior_pop_colors_mag} is $p(\pmb{C})\rightarrow p(\pmb{C}|m_i)$. We compute  $p(\pmb{C}|m_i)$ by binning galaxies in 4 magnitude bins with equal numbers, and linearly interpolating the values of those four bins to any value of $m_i$.

\begin{figure}
	\includegraphics[width=\columnwidth]{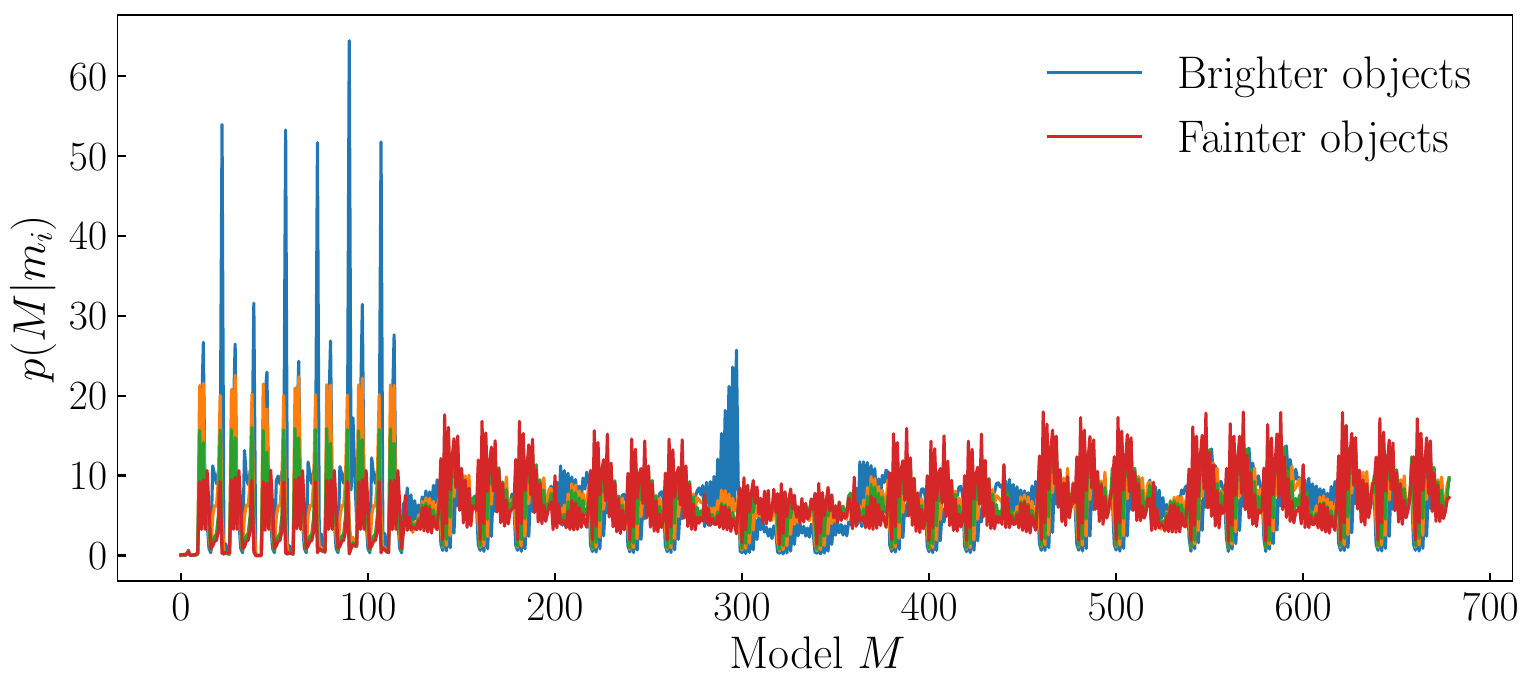}
    \caption{The prior for each model $M$ used in this work for each magnitude bin, $p(M|m_i)$. There are four magnitude bins with equal number of galaxies for which $p(M|m_i)$ is computed, which appear color coded from brighter to fainter as: blue, orange, green and red. The prior is computed using the observed colors of the data, and the colors spanned by each model $M$. See Appendix~\ref{app:popprior} and section~\ref{sec:population_prior} for details.}
    \label{fig:priormodel}
\end{figure}

Figure~\ref{fig:priormodel} shows the model prior used in this work for the four magnitude bins, $p(M|m_i)$. There are a total of 679 models in this work, featuring a number of different continuum templates groups, extinction laws and emission line templates. The figure shows the models ordered following Table~\ref{tab:bayesev_models}, putting models with the same continuum group but different emission line models or extinction law together.

\bsp	
\label{lastpage}
\end{document}